\documentclass{ieeeaccess}
\usepackage{cite}
\usepackage{amsmath,amssymb,amsfonts}
\usepackage{algorithmic}
\usepackage[ruled,vlined]{algorithm2e}
\usepackage{graphicx}
\usepackage{textcomp}
\usepackage[nolist]{acronym}
\usepackage{booktabs}
\usepackage{float}
\usepackage{stfloats}
\usepackage{fontawesome}
\usepackage{lmodern}
\usepackage{booktabs} 
\usepackage{tabularx} 

\newcommand{\Yes}{\faCircle}   
\newcommand{\No}{\faCircleO}   
\usepackage{hyperref} 
\hypersetup{
    colorlinks=true, 
    linkcolor=black, 
    urlcolor=black,   
    citecolor=black   
}

\usepackage{bm}
\makeatletter
\AtBeginDocument{\DeclareMathVersion{bold}
\SetSymbolFont{operators}{bold}{T1}{times}{b}{n}
\SetSymbolFont{NewLetters}{bold}{T1}{times}{b}{it}
\SetMathAlphabet{\mathrm}{bold}{T1}{times}{b}{n}
\SetMathAlphabet{\mathit}{bold}{T1}{times}{b}{it}
\SetMathAlphabet{\mathbf}{bold}{T1}{times}{b}{n}
\SetMathAlphabet{\mathtt}{bold}{OT1}{pcr}{b}{n}
\SetSymbolFont{symbols}{bold}{OMS}{cmsy}{b}{n}
\renewcommand\boldmath{\@nomath\boldmath\mathversion{bold}}}
\makeatother

\def\BibTeX{{\rm B\kern-.05em{\sc i\kern-.025em b}\kern-.08em
    T\kern-.1667em\lower.7ex\hbox{E}\kern-.125emX}}

\begin{document}

\history{Date of publication xxxx 00, 0000, date of current version xxxx 00, 0000.}
\doi{10.1109/ACCESS.2024.0429000}

\title{AGORA: Agentic Green Orchestration Architecture for Beyond 5G Networks}

\author{\uppercase{Rodrigo Moreira}\authorrefmark{1}\IEEEmembership{Member, IEEE}, \uppercase{Larissa Ferreira {Rodrigues Moreira}}\authorrefmark{1}\IEEEmembership{Member, IEEE}, 
\uppercase{Maycon Peixoto}\authorrefmark{2}, and \uppercase{Flávio {de Oliveira Silva}}\authorrefmark{3} \IEEEmembership{Senior Member, IEEE}}

\address[1]{Institute of Exact and Technological Sciences - Federal University of Viçosa (UFV), Brazil (e-mail: rodrigo@ufv.br, larissa.f.rodrigues@ufv.br)}
\address[2]{Institute of Computing - Federal University of Bahia (UFBA), Salvador, Brazil (e-mail: maycon.leone@ufba.br)}
\address[3]{Department of Informatics - ALGORITMI Centre - University of Minho (UMinho), Portugal (e-mail: flavio@di.uminho.pt)}

\tfootnote{This study was financed in part by the Coordenação de Aperfeiçoamento de Pessoal de Nível Superior – Brasil (CAPES) – Finance Code 001. Rodrigo Moreira gratefully acknowledges the financial support of FAPEMIG (Grant \#APQ00923-24). We also acknowledge the financial support of the FAPESP MCTIC/CGI Research project 2018/23097-3 - SFI2 - Slicing Future Internet Infrastructures and Fundação para a Ciência e Tecnologia (FCT) within the RD Units Project of Centro ALGORITMI.}

\markboth
{Moreira \headeretal: AGORA: Agentic Green Orchestration Architecture for Beyond 5G Networks}
{Moreira \headeretal: AGORA: Agentic Green Orchestration Architecture for Beyond 5G Networks}

\corresp{Corresponding author: Rodrigo Moreira (e-mail: rodrigo@ufv.br).}

\acrodef{3GPP}{3rd Generation Partnership Project}
\acrodef{5G}{5th Generation Mobile Network}
\acrodef{6G}{6th Generation Mobile Network}
\acrodef{AI}{Artificial Intelligence}
\acrodef{API}{Application Programming Interface}
\acrodef{AI4Net}{\ac{AI} for Networking}
\acrodef{AIDER}{Aerial Image Dataset for Emergency Response}
\acrodef{AMF}{Access and Mobility Management Function}
\acrodef{AIaaS}{Artificial Intelligence-as-a-Service}
\acrodef{AGORA}{Agentic Green Orchestration Architecture for Beyond 5G Networks}
\acrodef{IID}{Independent and Identically Distributed}
\acrodef{B5G}{Beyond Fifth Generation}
\acrodef{BPF}{Berkeley Packet Filter}
\acrodef{CBR}{Constant Bit Rate}
\acrodef{CSV}{Comma-Separated Values}
\acrodef{CPU}{Central Processing Unit}
\acrodef{CNN}{Convolutional Neural Network}
\acrodef{CNNs}{Convolutional Neural Networks}
\acrodef{DoS}{Denial of Service}
\acrodef{DDoS}{Distributed Denial of Service}
\acrodef{DDPG}{Deep Deterministic Policy Gradient}
\acrodef{DNN}{Deep Neural Network}
\acrodef{DRL}{Deep Reinforcement Learning}
\acrodef{DT}{Decision Tree}

\acrodef{ETSI}{European Telecommunications Standards Institute}
\acrodef{eNWDAF}{Evolved Network Data Analytics Function}
\acrodef{EIF}{Energy Information Function}
\acrodef{E2E}{End-to-End}
\acrodef{FIBRE}{Future Internet Brazilian Environment for Experimentation}
\acrodef{FPR}{False Positive Rate}
\acrodef{GNN}{Graph Neural Networks}
\acrodef{GPU}{Graphics Processing Unit}
\acrodef{GTP}{GPRS Tunnelling Protocol}
\acrodef{HTM}{Hierarchical Temporal Memory}

\acrodef{IAM}{Identity And Access Management}
\acrodef{ICMP}{Internet Control Message Protocol}
\acrodef{IID}{Informally, Identically Distributed}
\acrodef{IoE}{Internet of Everything}
\acrodef{IoT}{Internet of Things}
\acrodef{ITU}{International Telecommunication Union}
\acrodef{IBN}{Intent-Based Networking}
\acrodef{IoA}{Internet of Agents}
\acrodef{KNN}{K-Nearest Neighbors}
\acrodef{KPI}{Key Performance Indicator}
\acrodef{KPIs}{Key Performance Indicators}
\acrodef{LSTM}{Long Short-Term Memory}
\acrodef{LLM}{Large Language Model}
\acrodef{MAE}{Mean Absolute Error}
\acrodef{ML}{Machine Learning}
\acrodef{MLaaS}{Machine Learning as a Service}
\acrodef{MOS}{Mean Opinion Score}
\acrodef{MAPE}{Mean Absolute Percentage Error}
\acrodef{MSE}{Mean Squared Error}
\acrodef{MEC}{Multi-access Edge Computing}
\acrodef{mMTC}{Massive Machine Type Communications}
\acrodef{MFA}{Multi-factor Authentication}
\acrodef{MCP}{Model Context Protocol}
\acrodef{NWDAF}{Network Data Analytics Function}
\acrodef{Net4AI}{Networking for \ac{AI}}
\acrodef{NS}{Network Slicing}
\acrodef{OSM}{Open Source MANO}
\acrodef{PCA}{Principal Component Analysis}
\acrodef{PoC}{Proof of Concept}
\acrodef{PPV}{Positive Predictive Value}
\acrodef{QoE}{Quality of experience}
\acrodef{QoS}{Quality of Service}
\acrodef{RAM}{Random Access Memory}
\acrodef{RF}{Random Forest}
\acrodef{RL}{Reinforcement Learning}
\acrodef{RMSE}{Root Mean Square Error}
\acrodef{RNN}{Recurrent Neural Network}
\acrodef{RTT}{Round-Trip Time}
\acrodef{RAN}{Radio Access Network}

\acrodef{SDN}{Software-Defined Networking}
\acrodef{SFI2}{Slicing Future Internet Infrastructures}
\acrodef{SLA}{Service-Level Agreement}
\acrodef{SON}{Self-Organizing Network}
\acrodef{SMF}{Session Management Function}
\acrodef{S-NSSAI}{Single Network Slice Selection Assistance Information}
\acrodef{SATSI}{Space-Air-Terrestrial-Sea Integrated Networks}

\acrodef{TQFL}{Trust Deep Q-learning Federated Learning}
\acrodef{TEID}{Tunnel Endpoint Identifier}
\acrodef{TPS}{Transactions Per Second}
\acrodef{TPR}{True Positive Rate}
\acrodef{UE}{User Equipment}
\acrodef{UEs}{User Equipments}
\acrodef{UPF}{User Plane Function}
\acrodef{UPFs}{User Plane Functions}
\acrodef{PDU}{Packet Data Unit}
\acrodef{URLLC}{Ultra-Reliable and Low Latency Communications}
\acrodef{UAV}{Unmanned Aerial Vehicle}
\acrodef{UAVs}{Unmanned Aerial Vehicles}
\acrodef{UDP}{User Datagram Protocol}
\acrodef{VoD}{Video on Demand}
\acrodef{VR}{Virtual Reality}
\acrodef{V2X}{Vehicle-to-Everything}

\acrodef{ZTN}{Zero-Touch Network}
\begin{abstract}
Effective management and operational decision-making for complex mobile network systems present significant challenges, particularly when addressing conflicting requirements such as efficiency, user satisfaction, and energy-efficient traffic steering. The literature presents various approaches aimed at enhancing network management, including the Zero-Touch Network (ZTN) and Self-Organizing Network (SON); however, these approaches often lack a practical and scalable mechanism to consider human sustainability goals as input, translate them into energy-aware operational policies, and enforce them at runtime. In this study, we address this gap by proposing the AGORA: Agentic Green Orchestration Architecture for Beyond 5G Networks. AGORA embeds a local tool-augmented Large Language Model (LLM) agent in the mobile network control loop to translate natural-language sustainability goals into telemetry-grounded actions, actuating the User Plane Function (UPF) to perform energy-aware traffic steering. The findings indicate a strong latency-energy coupling in tool-driven control loops and demonstrate that compact models can achieve a low energy footprint while still facilitating correct policy execution, including non-zero migration behavior under stressed Multi-access Edge Computing (MEC) conditions. Our approach paves the way for sustainability-first, intent-driven network operations that align human objectives with executable orchestration in Beyond-5G infrastructures.
\end{abstract}


\titlepgskip=-21pt

\maketitle

\section{Introduction}\label{sec:introduction}

Future mobile network architectures are required to deliver cognitive services to end users while simultaneously ensuring sustainable network management~\cite{Trevlakis2024, Umut2025}. The provision of cognitive services allows users to utilize \ac{AI} capabilities according to their needs at the network edge~\cite{RodriguesMoreira2023, Larissa2024, Moreira2025}. Although \ac{B5G} necessitates complex closed-loop systems for management and orchestration, it will drive innovation across industries and smart societies by offering not only unprecedented network \acp{KPI} capabilities but also human-centered features and applications~\cite{Bello2024, Sun2025}. This evolution presents challenges for management and operations across various network segments, particularly as modern mobile networks incorporate \ac{SATSI} and facilitate the Internet of Smart Things~\cite{pivoto2023, Xiao2024}. Nevertheless, \ac{AI} has enhanced the robustness of network management in addressing complex, multi-operator, and vendor-related challenges. Yet, turning high-level operational goals, especially sustainability goals, into \ac{AI}-native, safe, measurable, and executable actions across the data plane remains a largely unresolved challenge~\cite{10.1007/978-3-031-28451-9_11, Bello2024, Sun2025}.

In the literature, we found different approaches to manage complex mobile network architectures, as expected in \ac{6G}. \ac{3GPP} standardized \ac{SON} mechanisms, and the field has since progressed toward \ac{ZTN} and \ac{IBN}, reshaping how network operations requirements are specified and automated~\cite{Jiang2024, Sedat2025, Alex2025, Dazzi2025, Mekrache2025, Cheng2025}. As autonomy increases, the control logic shifts from static policies toward distributed agent ecosystems. More recently, a new paradigm has been emerging: the \ac{IoA}. The \ac{IoA} establishes a distributed framework in which autonomous, \ac{AI}-driven agents interconnect through standardized protocols to collaborate, exchange knowledge, and orchestrate workflows across decentralized environments~\cite{Acharya2025, Wang_2025}. In this context, \ac{IoA} refers to the interconnected multi-agent ecosystem, whereas agentic characterizes the autonomous reasoning and goal-driven behavior exhibited by individual agents within that ecosystem. A common thread across these approaches is the pursuit of higher autonomy through standardized closed loops and intent-based specifications~\cite{Njah2025}. However, sustainability objectives are often treated as secondary constraints, and they rarely translate into concrete, telemetry-grounded control actions at the \ac{5G} core data plane.

\textbf{Problem statement.} Current intent-based and zero-touch frameworks lack a practical mechanism to translate human sustainability goals into verifiable, tool-executed \ac{UPF} actions that are continuously grounded in real-time telemetry.

This gap becomes more pronounced in Beyond \ac{5G} deployments, where decisions must span heterogeneous domains, including the data plane, \ac{MEC}, and the \ac{5G} core, under time-varying loads. Leveraging an \ac{LLM}, the agentic framework can perform reasoning and planning, use tools, manage memory, and collaborate with other agents, thereby enhancing network management with unprecedented capabilities. To the best of our knowledge, existing approaches do not directly actuate the \ac{5G} \ac{UPF} to achieve green network management objectives. Consequently, this study proposes AGORA: Agentic Green Orchestration Architecture for Beyond \ac{5G} Networks, which integrates the local deployment of \acp{LLM} into data plane management. This integration aims to translate aspirations for greener network management into complex network management tasks, thereby augmenting the established \ac{ZTN} or \ac{SON} management approaches. Building on this design, we implemented an end-to-end prototype. We evaluated it under controlled \ac{MEC} stress to quantify the energy footprint, policy compliance, and \ac{UE} perceived \ac{QoS}.

The main contributions of this paper are as follows: (i) an agentic \ac{LLM} driven closed-loop architecture that translates natural language intents into telemetry-grounded tool calls and \ac{UPF} routing actions for sustainable \ac{B5G} management; (ii) a policy compliance evaluation methodology for tool-augmented agents, including an intent suite with phrasing variations, tool and action compliance indicators, and energy and \ac{UE} \ac{QoS} metrics aligned to decision windows; (iii) an end-to-end prototype to create controlled \ac{MEC} energy asymmetry; and (iv) an experimental comparison of multiple local \acp{LLM}, including a non-English capable model, quantifying energy footprint, latency, and migration behavior under an energy threshold policy. Accordingly, our evaluation focuses on whether AGORA closes this gap by (i) grounding decisions on telemetry via tool calls and (ii) enforcing the resulting green policy directly at the \ac{UPF} under \ac{MEC} stress.

The remainder of this paper is organized as follows: Section~\ref{sec:related_work} contrasts our approach with those in the literature, while Section~\ref{sec:proposed_method} presents our method. Section~\ref{sec:experimental_testbed} describes our evaluation testbed, and Section~\ref{sec:results_and_dicussion} presents the results and discussions of our approach. Finally, we present concluding remarks and possible research directions in Section~\ref{sec:concluding_remarks}.

\section{Related Work}\label{sec:related_work}

The orchestration of \ac{B5G} and \ac{6G} networks has recently transitioned toward intent-based and autonomous paradigms, driven by \acp{LLM} and agentic architectures. This section categorizes recent advancements into objective-driven orchestration, intent-to-action translation, and agentic control loops.

\subsection{Zero-Touch and Objective-Driven Orchestration}
Early efforts in green edge computing established the foundation for objective-driven frameworks. Guim et al. \cite{Guim2022} introduced a multi-tier lifecycle management system using Kubernetes Custom Resource Definitions (CRDs) to translate performance targets into scaling actions. Similarly, Barrachina Muñoz et al. \cite{Munoz2024} operationalized Zero-Touch Management (ZSM) through the MonB5G architecture, employing a Monitor-Analyze-Decide-Execute (MAPE) loop for slice reconfiguration. While these works facilitate automated management, they primarily rely on deterministic policies or classical \ac{ML} and lack the flexibility of \ac{LLM}-based reasoning in complex, multi-domain environments. Decentralized approaches, such as the SCHE2MA framework \cite{Dalgkitsis2023}, utilize \ac{RL} to balance latency and energy consumption, yet often treat resource descriptors and reward weights as opaque parameters.

\subsection{LLM-Enabled Intent Translation and OSS Integration}
A significant body of work explores LLMs as intermediaries between natural language intents and network configurations. Dandoush et al. \cite{Dandoush2024} and Tzanakaki et al. \cite{Tzanakaki2025} proposed hierarchical agent workflows and transformer-based translators to map user requirements into standards-based descriptors, such as Open Source MANO (OSM) and YANG templates. In the realm of Operations Support Systems (OSS), Mekrache et al. \cite{Mekrache2025, OSSGPT2025} developed GPT-based interfaces that decompose intents into ordered API calls across heterogeneous domains. While effective for provisioning, these approaches prioritize the initial deployment and intent reconciliation as seen in the MAESTRO \cite{Maestro2024} and AIORA \cite{Molner2025} architectures rather than continuous, telemetry-grounded closed-loop control.

\subsection{Agentic Control Loops and Sustainability}
The latest frontier involves ``agentic'' orchestration, where LLMs autonomously invoke tools to manage live infrastructure. Recent frameworks like Brodimas et al. \cite{Brodimas2025} and Elkael et al. \cite{Elkael2025} leverage tool-calling mechanisms and the \ac{MCP} to execute Kubernetes and protocol-stack commands. Chatzistefanidis et al. \cite{Chatzistefanidis2025, MXAI2025, AGORAN2025} further extended this to Open RAN environments, using agent graphs to enforce blueprints via R1/E2 interfaces. 

Although these works introduce sophisticated reasoning, sustainability metrics, such as real-time carbon intensity or power efficiency, are often treated as secondary constraints or omitted entirely. Notable exceptions include Habib et al. \cite{Habi2025, HabibLLM2025} and Chergui et al. \cite{Chergui2025}, who integrate \ac{RL}, Digital Twins, and risk metrics to optimize power consumption; however, their focus remains mainly on Radio Access Network (RAN) scheduling rather than \ac{E2E} sustainability orchestration.

\subsection{Research Gap}

\begin{table*}[ht]
\centering
\caption{Comparison of related approaches across key capabilities.}
\label{tab:related-approaches}
\resizebox{\textwidth}{!}{%
\begin{tabular}{@{}cccccc@{}}
\toprule
\textbf{Approach}                         & \textbf{LLM Agentic} & \textbf{Native Tool-Calling} & \textbf{Telemetry Feedback} & \textbf{E2E B5G Scope} & \textbf{Sustainability Focus} \\ \midrule
Guim et al.~\cite{Guim2022}               & \No                   & \No                           & \Yes                         & \No                     & \Yes                           \\
Dalgkitsis et al.~\cite{Dalgkitsis2023}   & \No                   & \No                           & \Yes                         & \No                     & \Yes                           \\
Dandoush et al.~\cite{Dandoush2024}       & \Yes                  & \No                           & \No                          & \Yes                    & \No                            \\
Maestro~\cite{Maestro2024}                & \Yes                  & \No                           & \No                          & \No                     & \No                            \\
MonB5G~\cite{Munoz2024}                   & \No                   & \No                           & \Yes                         & \Yes                    & \No                            \\
Tzanakaki et al.~\cite{Tzanakaki2025}     & \No                   & \No                           & \No                          & \Yes                    & \No                            \\
DMO-GPT~\cite{Mekrache2025}               & \Yes                  & \Yes                          & \No                          & \Yes                    & \No                            \\
OSS-GPT~\cite{OSSGPT2025}                 & \Yes                  & \Yes                          & \No                          & \No                     & \No                            \\
Brodimas et al.~\cite{Brodimas2025}       & \Yes                  & \Yes                          & \No                          & \No                     & \No                            \\
Symbiotic~\cite{Chatzistefanidis2025}     & \Yes                  & \No                           & \Yes                         & \No                     & \No                            \\
MX-AI~\cite{MXAI2025}                     & \Yes                  & \Yes                          & \Yes                         & \No                     & \No                            \\
AGORAN~\cite{AGORAN2025}                  & \Yes                  & \No                           & \No                          & \No                     & \No                            \\
AgentRAN~\cite{Elkael2025}                & \Yes                  & \Yes                          & \Yes                         & \No                     & \No                            \\
AIORA~\cite{Molner2025}                   & \No                   & \No                           & \No                          & \Yes                    & \No                            \\
Habib et al.~\cite{Habi2025,HabibLLM2025} & \No                   & \No                           & \Yes                         & \No                     & \Yes                           \\
Chergui et al.~\cite{Chergui2025}         & \Yes                  & \No                           & \No                          & \No                     & \No                            \\ \midrule
\textbf{Our Proposal}                     & \textbf{\Yes}         & \textbf{\Yes}                 & \textbf{\Yes}                & \textbf{\Yes}           & \textbf{\Yes}                  \\ \bottomrule
\end{tabular}%
}
\end{table*}

Table \ref{tab:related-approaches} presents a comparative analysis of the proposed architecture with the current state-of-the-art architecture, evaluated across five critical technical pillars. We denote ({\tiny \faCircle}) when the approach achieves the feature and ({\tiny \faCircleO}) when it does not. The ``LLM Agentic'' column distinguishes frameworks that employ autonomous reasoning agents from those restricted to passive intent translation or deterministic logic. The ``Native Tool-Calling'' criterion identifies systems capable of direct, programmatic interaction with live infrastructure \acp{API} via function calls, rather than merely generating static deployment descriptors. The ``Telemetry Feedback'' criterion evaluates the presence of a dynamic closed-loop integration that utilizes real-time time-series data to inform agentic decision-making. Additionally, the ``E2E B5G Scope'' assesses whether the orchestration encompasses the entire Beyond-\ac{5G} landscape, including Core and \ac{MEC} domains, rather than being confined to isolated segments such as the \ac{RAN}. Finally, the ``Sustainability Focus'' indicates whether energy, power, or carbon metrics are prioritized as primary orchestration objectives rather than secondary constraints. 

 In contrast to the cited literature, our approach unifies real-time observability with an agentic tool-calling framework specifically optimized for \ac{E2E} sustainability. While prior efforts focus on intent-to-descriptor translation or narrow-domain closed loops, our work operationalizes an \ac{LLM} agent that ingests high-fidelity telemetry and fault-injection signals to autonomously call functions across Kubernetes, \ac{5G} Core, and \ac{MEC} interfaces. This positioning ensures that power, energy, and carbon efficiency are treated as first-class objectives guiding the automated re-orchestration process.

\section{Proposed Method}\label{sec:proposed_method}

\begin{figure*}[htpb]
  \centering
  \includegraphics[width=0.8\textwidth]{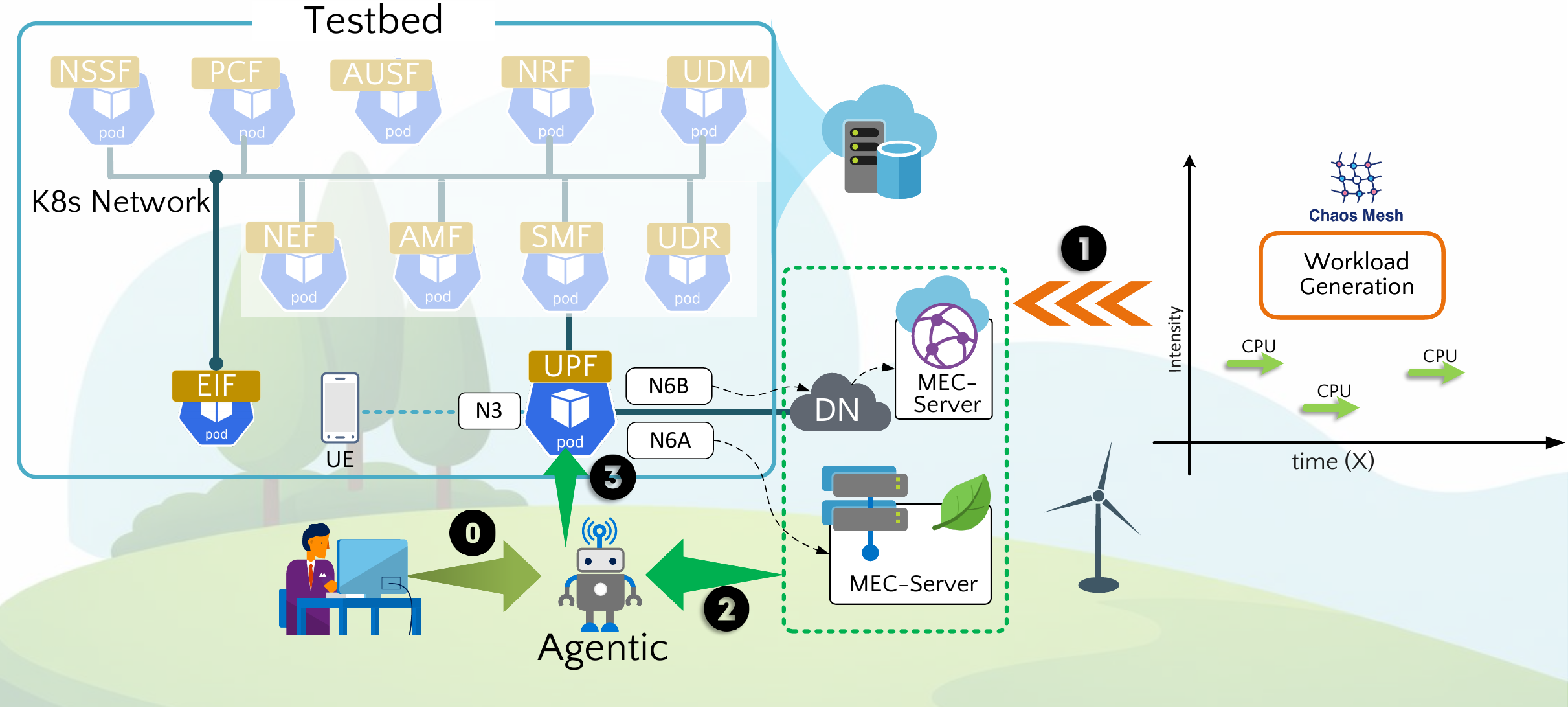}
  \caption{Agentic closed-loop sustainability orchestration in a 5G MEC testbed.}
  \label{fig:method}
\end{figure*}

Mobile networks increasingly require energy-aware closed-loop orchestration that can translate high-level operational goals into concrete, real-time actions across heterogeneous infrastructures. Figure~\ref{fig:method} summarizes AGORA, the end-to-end workflow of our sustainable agentic orchestration in the 5G testbed. In Phase (0), the user expresses a high-level sustainability intent that defines a green policy goal, according to Table~\ref{tab:instructions}. In Phase (1), a controlled workload is injected into the \ac{MEC} environment using a chaos tool to emulate realistic~\cite{Pedroso2025}, time-varying operating conditions and create energy asymmetry across \ac{MEC} sites, where MEC2 is \ac{GPU}-enabled for cognitive services and MEC1 is \ac{CPU}-only as the greener target. In Phase (2), the agentic controller queries the monitoring stack to estimate the current system state, including the power and performance signals. In Phase (3), the agent enforces the policy by actuating the 5G data plane at the UPF and updating the routing rules toward the greener \ac{MEC} target based on the observed telemetry.

\subsection{Overview and Research Goal}
\label{subsec:method_overview}

This work evaluates whether an agentic with local \ac{LLM} can act as a network manager in a \ac{B5G} control loop, where correctness is defined by the ability to (i) interpret user intents through internal reasoning, (ii) select and invoke the appropriate external tools, and (iii) enforce an energy policy by applying a greener routing action at the \ac{UPF}. The method isolates the decision-making capability of tool-augmented agents from the experimental testbed details, which are reported separately.

Our method specifies the decision loop, the tool interfaces, the intent suite, the compliance criteria, and the derived metrics. Implementation and deployment details such as cluster configuration, \ac{UPF} realization, telemetry stack versions, and hardware resources are described in the experimental testbed section to avoid conflating algorithmic design with platform-specific factors.

\subsection{Agentic Control Loop}\label{subsec:method_agentic_loop}

We model network management as a closed loop with four stages: intent ingestion, tool-based state estimation, policy evaluation, and actuation. Let $t$ be a decision step triggered by a user request. The agent receives a natural language intent $I_t$, queries measurements through tools to obtain an observed state $\hat{s}_t$, evaluates a policy $\pi(\hat{s}_t, I_t)$, and issues an action $a_t$ that updates the \ac{UPF} target.

The actuation space is discrete and corresponds to selecting the serving \ac{MEC} site (according to Equation~\ref{eq:1}):
\begin{equation}
\label{eq:1}
a_t \in \{ \texttt{route\_to\_MEC1}, \texttt{route\_to\_MEC2} \}.
\end{equation}

Our method is implemented with a chat-oriented \ac{LLM} bound to two executable tools:
\begin{itemize}
  \item \texttt{energy\_mean\_last\_time(mec, delta)} returns an energy proxy for the requested MEC, derived from infrastructure telemetry.
  \item \texttt{upf\_set\_target(mec)} updates the \ac{UPF} egress selection so that traffic is routed toward the chosen MEC.
\end{itemize}

In standards-compliant deployments, the agentic tool can be backed by energy telemetry exposed by the \ac{3GPP} \ac{EIF}~\cite{3gpp.23.501}, ensuring that power/energy \acp{KPI} are obtained via the management plane rather than ad hoc instrumentation.

To enable consistent cross-log analysis, each intent execution is associated with a decision interval $\Delta_t = [t^{\text{start}}_t, t^{\text{end}}_t]$ covering the \ac{LLM} inference and any tool invocations. All infrastructure telemetry and \ac{UE} probing samples are aligned to $\Delta_t$ by timestamp, using the nearest-neighbor matching when sampling periods differ.

\subsection{Intent Suite and Classification}
\label{subsec:method_intents}

To probe the robustness of energy-aware tool selection via AGORA reasoning and policy enforcement, we used a compact suite of user intents that varied in phrasing, urgency, and constraint expression. Each intent encodes a threshold-based policy and requires the agent to consult measurements before making a decision. The four intents used in the evaluation are listed in Table~\ref{tab:instructions}.

\begin{table}[ht]
\centering
\caption{User prompt phrasing variations for energy-aware traffic migration.}
\label{tab:instructions}
\small
\renewcommand{\arraystretch}{1.4} 
\begin{tabularx}{\columnwidth}{l X} 
\toprule
\textbf{Type} & \textbf{User Instruction Prompt.} \\ \midrule
Threshold & \textit{Monitor MEC2 power. If $> \theta$, migrate to MEC1.} \\ 
Policy-based & \textit{Is there any green policy violation on MEC2? If energy $> 20$\,W, trigger migration.} \\ 
Contextual & \textit{Compare MEC power and move traffic if MEC2 usage is too high.} \\ 
Urgent &\textit{ Execute migration to MEC1 now if MEC2 usage $> 20$\,W.} \\ \bottomrule
\end{tabularx}
\end{table}

We denote by $\theta(I_t)$ the threshold implicitly specified by the intent text. In our suite, $\theta(I_t)$ equals $\theta$ for the baseline threshold intent and equals $20$ W (empirically defined in AGORA as a power threshold) for the policy-based and urgent variants. This allows the same compliance rule to be evaluated across heterogeneous phrasings.

AGORA can handle non-English energy-aware prompts, such as those in Portuguese, making our method scalable for different \ac{LLM} models. This study aimed to evaluate whether higher-level user phrasing in English and Portuguese can still be mapped to correct tool usage and green-compliant actions.

\subsection{Workload Induction and Energy Asymmetry}
\label{subsec:method_workload}

The evaluation requires measurable, repeatable differences in energy consumption across \ac{MEC} sites. Therefore, we induced asymmetric load episodes over \ac{MEC} workloads, producing stress intervals with timestamps. The stress generator applies a randomized \ac{CPU} load with different parameter ranges per \ac{MEC}, yielding a sustained higher load on MEC2 and lighter or shorter stress on MEC1. Each stress event ($e_i$) was logged according to Equation~\ref{eq:2}.
\begin{equation}
\label{eq:2}
e_i = \langle \texttt{MEC}_{i}, \texttt{begin}, \texttt{end}, \texttt{cpu\_load}, \texttt{workers} \rangle
\end{equation}

The resulting stress log is later time-aligned with the agent decision traces to interpret behavior under stressed operating conditions.

This procedure does not assume that AGORA has direct access to the stress schedules. Instead, the agent must infer the operational condition from telemetry and reasoning insights obtained via tools and monitoring platforms, or even \ac{3GPP} \ac{EIF}.

\subsection{Policy and Decision Compliance}
\label{subsec:method_policy}

An energy-aware policy defines the target behavior. Let $\hat{P}_2(t)$ be the observed power proxy for MEC2 obtained by the tool at decision step $t$, and $\theta$ be the threshold specified by the intent. The policy compliance rule is given by Equation~\ref{eq:3}.

\begin{equation}
\centering
\label{eq:3}
a_t =
\begin{cases}
\texttt{route\_to\_MEC1} & \text{if } \hat{P}_2(t) > \theta, \\
\texttt{route\_to\_MEC2} & \text{otherwise.}
\end{cases}
\end{equation}

Because intent phrasing can be ambiguous, we evaluate compliance at two levels:
\begin{itemize}
  \item Tool compliance, whether the agent invokes energy measurement tools before acting.
  \item Action compliance, whether the resulting \ac{UPF} target matches the threshold-based rule implied by the intent.
\end{itemize}

An execution is considered valid only if the agent completes the loop in the order of state estimation, followed by actuation. Actuation without prior measurement is considered noncompliant, even if the final action coincidentally matches the threshold rule.

\subsection{Quality of Service Observation at the User Equipment}
\label{subsec:method_qos}

To quantify the user-perceived impact of AGORA actions, the \ac{UE} continuously probes the active (MEC$_{i}$) target selected by the \ac{UPF}. Each probing window logs the round-trip latency and \ac{UDP} quality indicators, producing records according to Equation~\ref{eq:4}.
\begin{equation}
\label{eq:4}
\centering
\begin{split}
u_j = \langle & t_s, t_e, \texttt{target\_MEC}, \texttt{ping\_avg\_ms}, \\
& \texttt{udp\_jitter\_ms}, \texttt{udp\_loss\_pct} \rangle.
\end{split}
\end{equation}

These measurements were later aligned with agent decisions to analyze how policy-driven migration affects perceived service quality.

Each \ac{UE} probing record $u_j$ is mapped to the most recent \ac{UPF} target observed at the start of its window, enabling the attribution of latency and \ac{UDP} quality to the active energy-aware routing choice during that time interval.

\subsection{Experimental Procedure for Model Comparison}
\label{subsec:method_procedure}

We evaluate AGORA with multiple \ac{LLM} backends as the energy-aware decision engine. For each model $M$, we repeat $R$ independent runs and execute the same intent suite $\mathcal{I}$ to estimate the average behavior under stochastic system conditions. As summarized in Algorithm~\ref{alg:agentic_eval}, each run initializes the agent and binds the external tools, then iterates over intents by (i) issuing the intent, (ii) querying telemetry via tool calls to estimate the current state, (iii) selecting the policy action, and (iv) actuating the \ac{UPF}. Finally, we persist the decision traces and align them with \ac{MEC} energy telemetry and \ac{UE} probing logs for cross-model aggregation.

\begin{algorithm}[h!]
\caption{Agentic policy evaluation per model.}
\label{alg:agentic_eval}
\DontPrintSemicolon
\KwIn{Model $M$, intent set $\mathcal{I}$, repetitions $R$, threshold $\theta$}
\For{$r \leftarrow 1$ \KwTo $R$}{
  Initialize the agent with model $M$ and bind tools\;
  \ForEach{$I_t \in \mathcal{I}$}{
    Provide $I_t$ to the agent\;
    Query telemetry via tool calls to obtain $\hat{P}_1(t)$ and $\hat{P}_2(t)$\;
    Apply the policy $\pi(\hat{P}_2(t), \theta)$ and select $a_t$\;
    Execute $a_t$ by calling \texttt{upf\_set\_target}\;
    Log tool calls, inference latency, token counts, and action outcome\;
  }
  Collect MEC energy telemetry snapshots aligned to the run\;
  Collect \ac{UE} probing records during the run window\;
}
Aggregate metrics across $R$ runs for model comparison\;
\end{algorithm}

\subsection{Measured Variables and Derived Metrics}
\label{subsec:method_metrics}

The method logs the decision time, token usage, tool call traces, and infrastructure energy proxies. For each AGORA intent execution, we recorded our local \ac{LLM} inference time $T_t$ and observed power proxy $\hat{P}_{m}(t)$ for the active \ac{MEC} $m$. We derive the energy consumed by the serving \ac{MEC} during decision execution using Equation~\ref{eq:5}.
\begin{equation}
\label{eq:5}
\centering
E^{\text{MEC}}_t = \hat{P}_{m}(t) \cdot T_t,
\end{equation}

Where $m$ equals MEC1 when migration is active; otherwise, it equals MEC2. We also compute the energy normalized by the output volume when applicable, according to Equation~\ref{eq:6}.

\begin{equation}
\label{eq:6}
\centering
E^{\text{token}}_t = \frac{E^{\text{gpu}}_t}{N^{\text{gen}}_t},
\end{equation}

Where $E^{\text{gpu}}_t$ is the GPU energy proxy during the decision interval and $N^{\text{gen}}_t$ is the number of generated tokens.

Finally, policy compliance is summarized using two binary indicators per intent execution, as defined in Equation~\ref{eq:7}.
\begin{equation}
\label{eq:7}
C^{\text{tool}}_t \in \{0,1\}, \quad C^{\text{act}}_t \in \{0,1\}.
\end{equation}
Here, $C^{\text{tool}}_t = 1$ if the AGORA invokes the telemetry measurement tools before acting, and $C^{\text{act}}_t = 1$ if the resulting \ac{UPF} target matches the threshold-based decision rule implied by the intent at step $t$. The resulting dataset enables a unified comparison of decision quality, energy footprint, and user-perceived \ac{QoS} impact across AGORA \ac{LLM} backends.

The above definitions specify the AGORA independently of the deployment. The experimental testbed section details the platform used to instantiate \ac{MEC} sites, collect telemetry, execute workload induction, and run \ac{UE} probing. In contrast, the present section defines how decisions, compliance, and energy and \ac{QoS} metrics are operationalized and compared across models.

\section{Experimental Testbed}\label{sec:experimental_testbed}

Here, we describe the testbed used to instantiate and evaluate AGORA in a container-based \ac{5G} core. We summarize the compute platform, the \ac{5G} core with two \ac{MEC} sites interconnected through a controllable \ac{UPF}, the local \ac{LLM} serving stack and evaluated models, and the monitoring instrumentation. We then describe the workload induction, \ac{UE} \ac{QoS} probing, and repeated run procedures used to ensure reproducible comparisons.

\subsection{Infrastructure and Compute Platform}\label{subsec:testbed_infra}

All experiments were conducted on the FABRIC testbed~\cite{fabric-2019} using an OpenStack-based compute node equipped with an AMD EPYC 7543 CPU and 64~GiB of system memory. The system exposes an NVIDIA A30 \ac{GPU} with 24GB of device memory for \ac{LLM} inference. The platform runs a Kubernetes cluster with both client and server versions at v1.28.15 that orchestrates the 5G core functions, \acp{MEC}, the monitoring stack, and workload components.

\subsection{5G Core and MEC Setup}\label{subsec:testbed_5g_mec}

We deploy free5GC, a \ac{5G} core, to provide end-to-end connectivity between the radio access and data networks. The user plane is anchored at a \ac{UPF} that exposes a simple control interface used by the agent to update the active routing target. Two MEC instances, denoted MEC1 and MEC2, are deployed as separate pods and are reachable from the \ac{UPF} through its data network interfaces. In the considered scenario, MEC1 represents the preferred green execution site, whereas MEC2 represents the stressed and non-green site used to evaluate policy compliance under adverse conditions. In our setup, MEC2 is \ac{GPU}-enabled to serve user cognitive workloads, whereas MEC1 is \ac{CPU}-only and represents the greener but less capable execution site.

\subsection{Agent Execution and LLM Serving}\label{subsec:testbed_llm_serving}

The AGORA decision engine is implemented as a tool-augmented chat \ac{LLM} that runs locally and issues actions to the network through external tools. Model inference was performed using vLLM with an OpenAI-compatible endpoint, enabling a consistent invocation interface across models and facilitating tool calling during inference. The serving configuration enables automatic tool selection and structured tool-call parsing, allowing the model to request telemetry and perform actuation within the same decision loop.

The evaluated models are:
\begin{itemize}
  \item Qwen2.5 1.5B Instruct~\cite{Qwen}, a compact instruction tuned model used to evaluate latency and tool calling behavior at a small parameter scale.
  \item Mistral 7B Instruct v0.2~\cite{Mixtral}, a mid-size instruction model used to represent a stronger general-purpose baseline.
  \item Phi 3.5 Mini Instruct~\cite{Phi3}, a small, efficient instruction model used to test the tradeoff between speed and decision compliance.
  \item OLMoE 1B 7B Instruct~\cite{OLMoE}, a mixture of experts style instruction model used to evaluate whether sparse activation improves energy proportionality.
  \item Sabi\'a 7B by Maritaca AI~\cite{maritaca}, a Portuguese capable model included to evaluate high-level intents in Portuguese and assess whether language alignment improves action compliance in green policies.
\end{itemize}

\subsection{Energy and Telemetry Instrumentation}\label{subsec:testbed_monitoring}
Infrastructure telemetry is collected through Kepler, which exposes container-level energy-related metrics to Prometheus. The monitoring stack continuously scrapes both the Kepler and vLLM metrics endpoints, enabling the synchronized collection of \ac{MEC} power proxies and inference-side counters, such as generated tokens, prompt tokens, and request inference time. All measurements were timestamped and later aligned with the agent decision windows to compute the derived energy metrics, such as active \ac{MEC} and \ac{GPU} energy proxies.

To ensure consistent access to telemetry during automated runs, Prometheus is accessed via a local port forward, and vLLM exposes its metrics and health endpoint on the local host. This configuration supports automated batch execution across repeated cold-start runs while maintaining a stable metric collection interface.

\subsection{Workload Induction and Stress Logging}\label{subsec:testbed_workload}
To create a measurable energy asymmetry between \ac{MEC} sites, we injected synthetic \ac{CPU} stress into \ac{MEC} pods using Chaos Mesh~\cite{chaosmesh}. The stress generator selects a target MEC and applies randomized load parameters, producing prolonged higher load intervals on MEC2 and shorter, lighter intervals on MEC1. Each stress episode is recorded in a structured log with start and end timestamps, load level, worker count, and duration. This log is used for post hoc contextualization of agent decisions, whereas the agent itself has no direct access to the stress schedule and must rely on telemetry obtained through the tool calls.

\subsection{User Equipment QoS Probing}\label{subsec:testbed_qos}

The user-perceived impact is measured at the \ac{UE} by periodic probing of the current \ac{UPF} routing target. The probing process repeatedly queries the target selection interface, extracts the active target IP address, and measures the round-trip time using \texttt{ping}. In addition, it optionally runs \ac{UDP} probing with \texttt{iperf3} to estimate the jitter and loss when available. All \ac{UE} records are timestamped and stored in a \ac{CSV} log, allowing alignment with the agent decisions and infrastructure telemetry during analysis.

\subsection{Experimental Execution Procedure}\label{subsec:testbed_procedure}

We induce controlled and repeatable load asymmetry across \ac{MEC} sites using Chaos Mesh by injecting \ac{CPU} stress episodes into the Kubernetes pods that host \ac{MEC} services. Each episode is defined by a target pod, \ac{CPU} load percentage, number of worker threads, and duration, and is applied as a \texttt{StressChaos} resource. The generator enforces a single active stressor at a time by cleaning up any existing \texttt{StressChaos} objects before scheduling the next episode. It logs every injected event, including begin and end timestamps, to a \ac{CSV} file. This produces a sustained higher load on MEC2 and a lighter load on MEC1, creating a consistent operating regime to evaluate whether agent decisions respond to measurable stress without exposing the stress schedule to the agent.

For each model, we executed multiple independent runs to capture the average behavior under stochastic system conditions and cold-start effects. At the beginning of each run, the vLLM server was started for the selected model, and the monitoring endpoints were verified via health checks. The agent is then prompted with the intent suite, and it may issue tool calls to query the \ac{MEC} power and update the \ac{UPF} target. After each run, logs are persisted, including the AGORA decision traces, \ac{MEC} energy \ac{CSV} snapshots, and \ac{UE} \ac{QoS} measurements. The vLLM process is terminated between runs to enforce cold-start conditions and reduce cross-run cache effects, enabling a fair comparison across models.

\section{Results and Discussion}\label{sec:results_and_dicussion}

Here, we discuss the AGORA test results for different local \acp{LLM}. We first examined how energy, delay, and \ac{UE} \ac{QoS} were affected by decision-making tools. Next, we examined how the workload context and \ac{GPU} behaved under stress.

\subsection{Energy-QoS Characterization of Tool-Augmented Agents}
\label{subsec:results_energy_qos}

We quantified (i) infrastructure impact at the edge and (ii) perceived \ac{UE} quality. Infrastructure energy is estimated by combining the Kepler instantaneous power at each \ac{MEC} with the agent inference duration per prompt. Unless otherwise stated, active \ac{MEC} energy denotes the energy spent at the \ac{MEC} that effectively serves the \ac{UPF} target during the agent decision loop, such as \ac{MEC}1 when migration is active; otherwise, \ac{MEC}2.

\begin{figure}[t]
  \centering
  \includegraphics[width=\columnwidth]{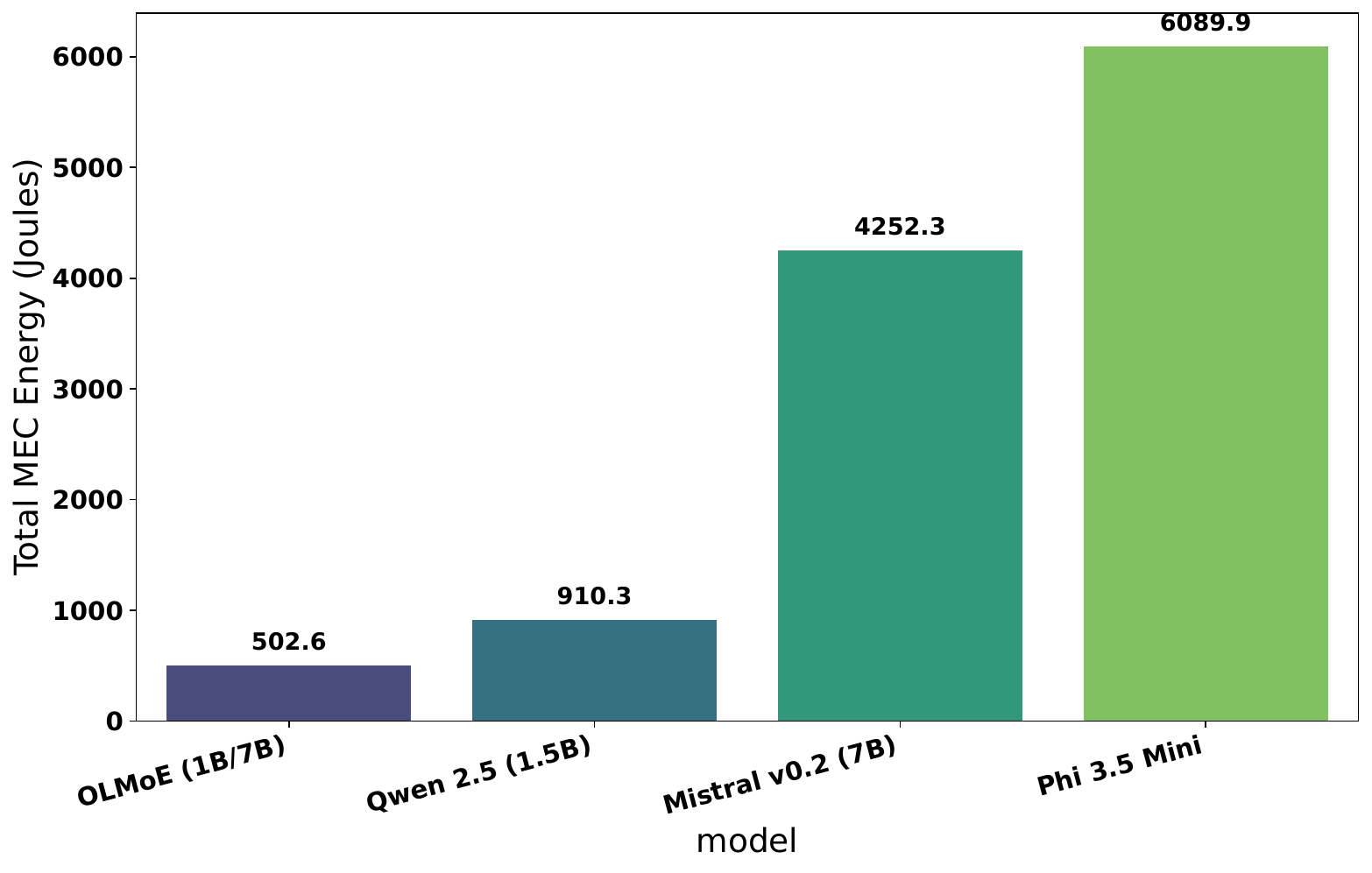}
  \caption{Total active \ac{MEC} energy across all runs.}
  \label{fig:total_mec_energy}
\end{figure}

\textbf{Aggregate infrastructure energy and latency}. Figure~\ref{fig:total_mec_energy} shows the total active \ac{MEC} energy for all runs. The difference is clear: OLMoE uses the least energy (502.6~J), followed by Qwen (910.3~J), while Mistral (4252.3~J) and Phi (6089.9~J) consume significantly more. This ranking demonstrates that, in this setting, the agent's runtime is the primary factor affecting the AGORA reasoning energy consumption, even when the \ac{MEC} is under stress.

\begin{figure}[t]
  \centering
  \includegraphics[width=\columnwidth]{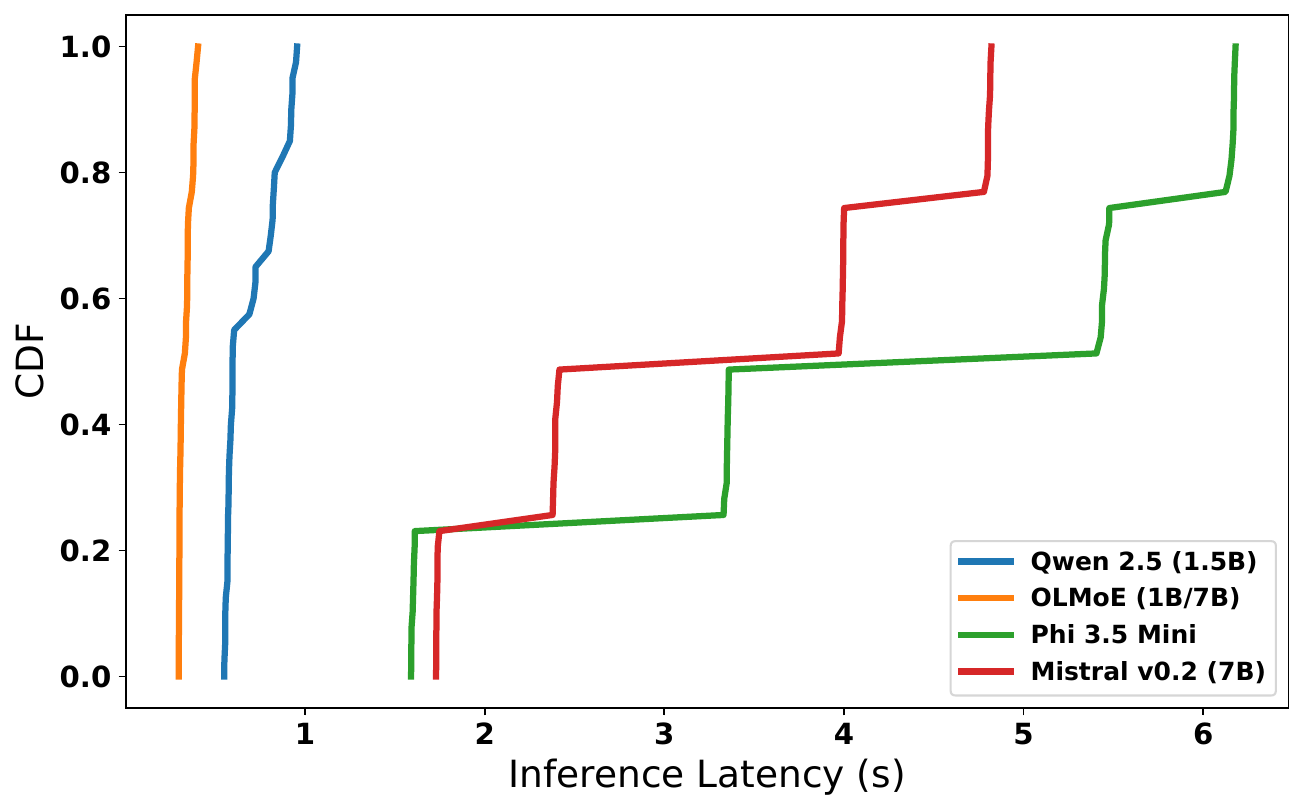}
  \caption{CDF of agent inference latency.}
  \label{fig:latency_cdf}
\end{figure}

The latency distribution in Figure~\ref{fig:latency_cdf} matches the energy ranking of the workloads. OLMoE was mostly under 0.6s. Qwen was between 0.7s and 1.0s. Mistral and Phi have longer times, with Phi exceeding 6s. These times are essential for understanding energy use. Longer decision times imply that the system uses more energy because the energy-greedy \ac{MEC} remains active for a longer duration.

\begin{figure}[t]
  \centering
  \includegraphics[width=\columnwidth]{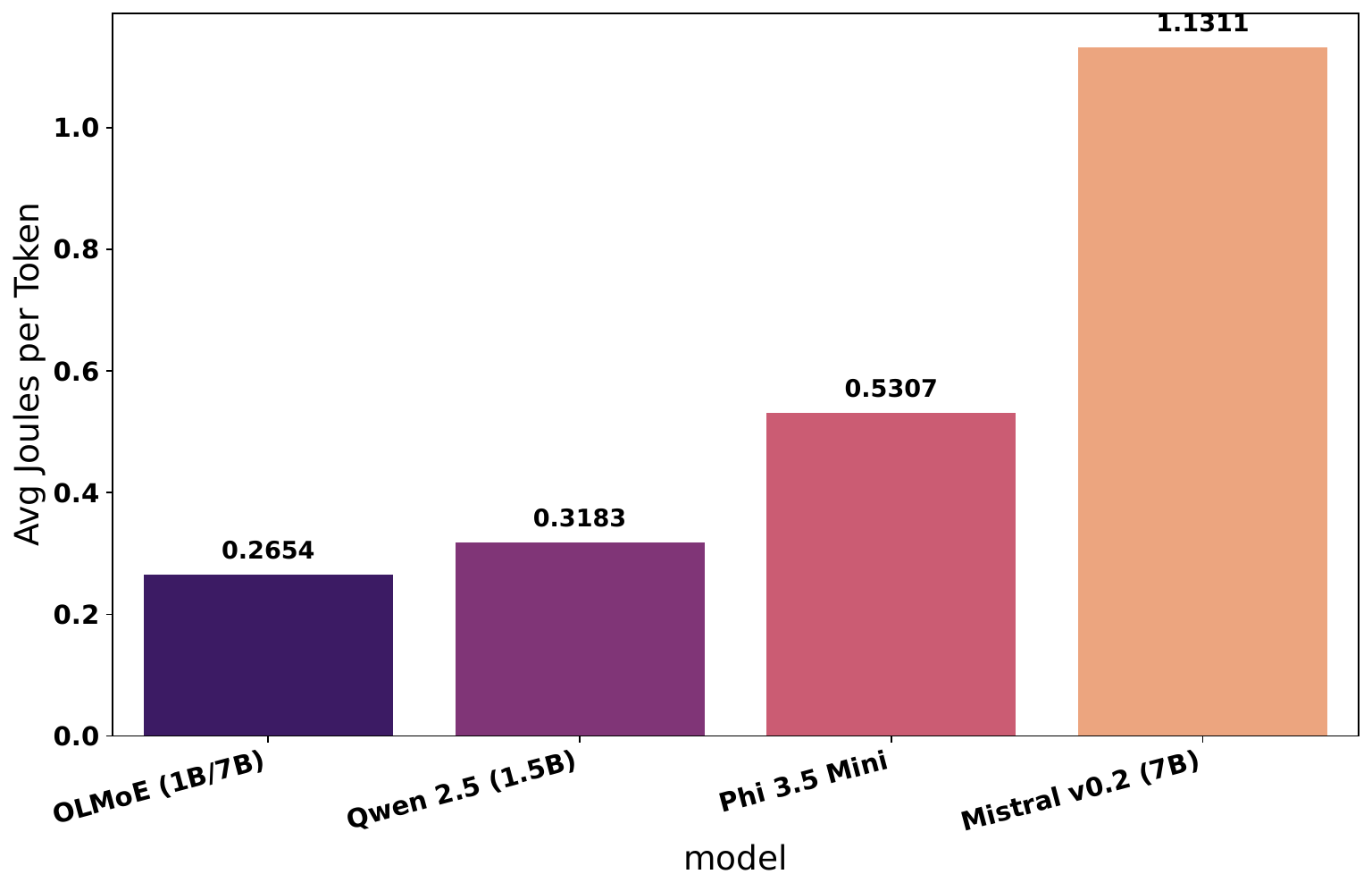}
  \caption{Average energy per generated token.}
  \label{fig:joules_per_token}
\end{figure}

\textbf{Energy efficiency normalized by output}. The total energy alone does not indicate whether a higher output volume compensates for longer runtimes. Figure~\ref{fig:joules_per_token} shows the normalized energy by the number of generated tokens. OLMoE remained the most efficient (0.2654~J/token), followed by Qwen (0.3183~J/token). Phi increases the energy per token to 0.5307~J/token, whereas Mistral is the least efficient (1.1311~J/token). This suggests that a proportionally greater token output does not offset the longer execution time of Mistral and that verbosity is an unreliable proxy for energy efficiency.

\begin{figure}[t]
  \centering
  \includegraphics[width=\columnwidth]{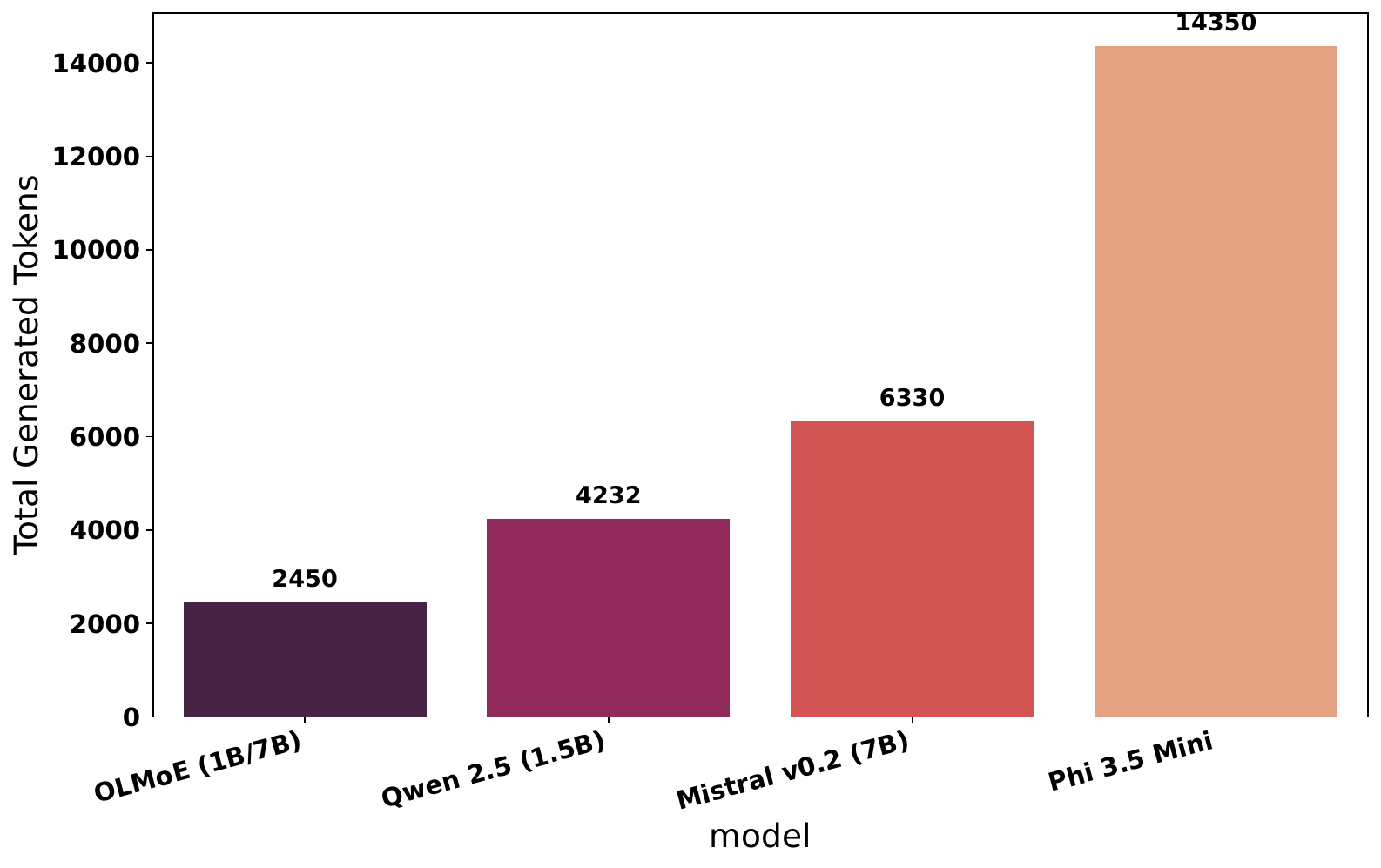}
  \caption{Total generated tokens across runs.}
  \label{fig:total_tokens}
\end{figure}

Figure~\ref{fig:total_tokens} contextualizes these findings by showing the total number of generated tokens. Phi produces the most tokens (14{,}350), followed by Mistral (6{,}330), Qwen (4{,}232), and OLMoE (2{,}450). When contrasted with Figure~\ref{fig:joules_per_token}, the results indicate an apparent decoupling between verbosity and efficiency; a model can generate many tokens while still being energy-inefficient per token.

\begin{figure}[t]
  \centering
  \includegraphics[width=\columnwidth]{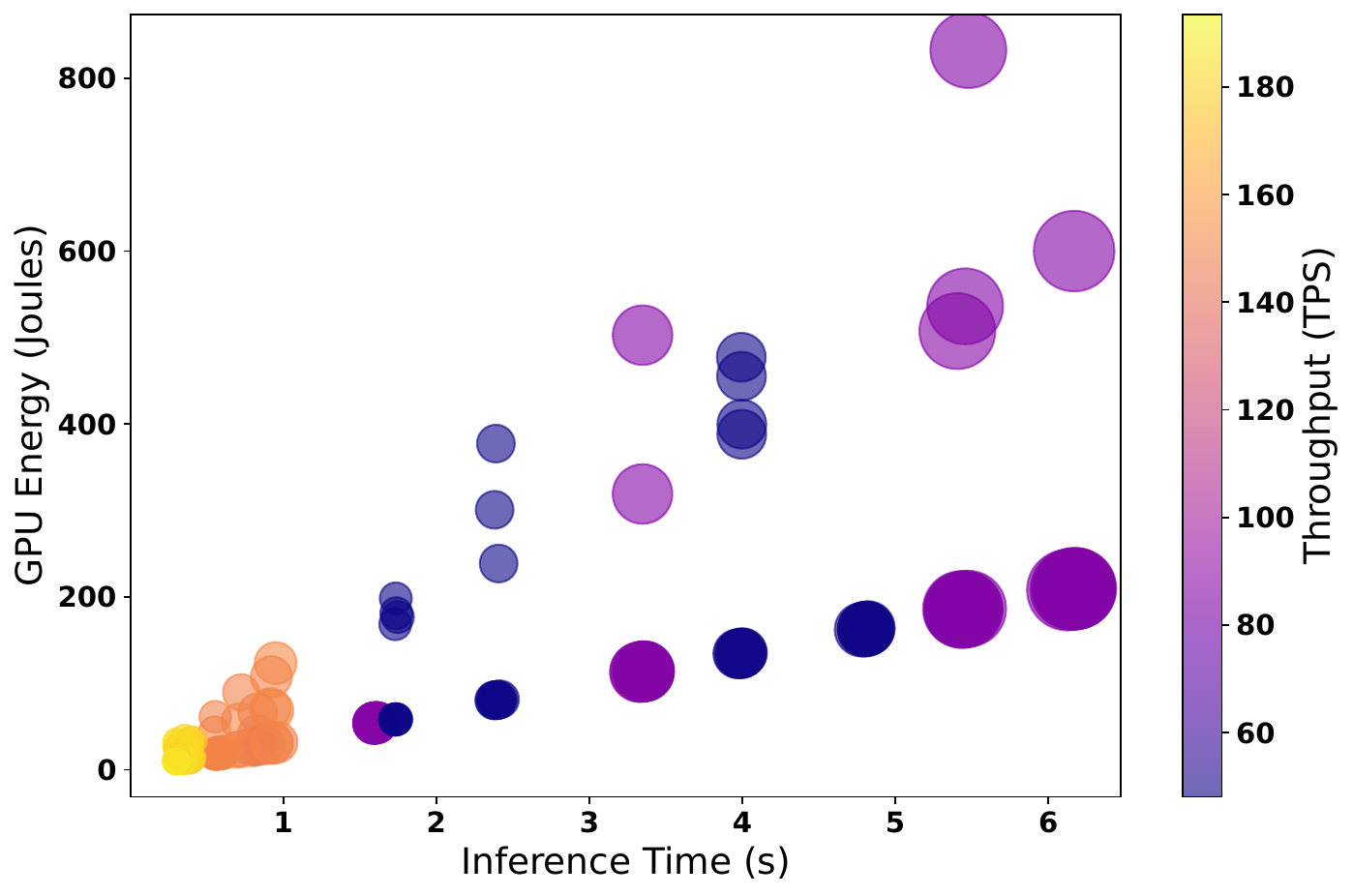}
  \caption{Inference time versus GPU energy with throughput and output volume.}
  \label{fig:efficiency_bubble}
\end{figure}

\textbf{System operating regimes}. Figure~\ref{fig:efficiency_bubble} combines the inference time (x-axis) with the GPU energy (y-axis), encoding the output volume via the marker size and the throughput via the color. The plot shows strong runtime-energy coupling, with the points moving upward as inference time increases. OLMoE and Qwen concentrate in high-throughput regions at shorter runtimes and lower GPU energy, whereas Mistral and Phi populate lower-throughput regimes, where longer runtimes amplify energy. This evidence supports the interpretation that, for tool-driven decision loops, faster models improve energy proportionality not only by reducing the time-to-decision but also by operating in higher-throughput regions of the accelerator.

\begin{figure}[t]
  \centering
  \includegraphics[width=\columnwidth]{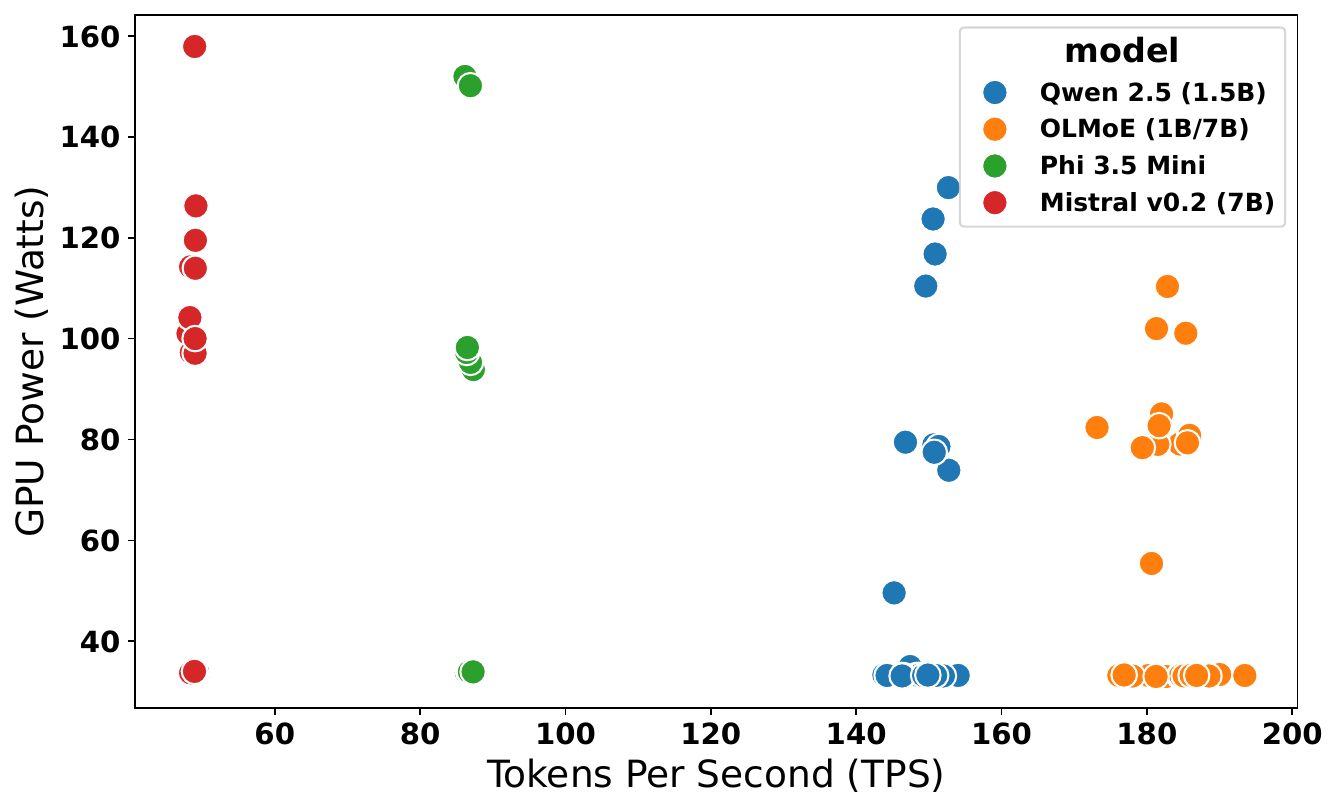}
  \caption{Throughput versus GPU power draw.}
  \label{fig:tps_vs_power}
\end{figure}

The relationship between the throughput and \ac{GPU} power is shown in Figure~\ref{fig:tps_vs_power}. OLMoE and Qwen achieved higher throughput (approximately 140–195 tokens/s) under moderate power conditions. In contrast, Mistral and Phi typically functioned below 90 tokens/s, yet they reached high power levels in several instances. Collectively, Figures~\ref{fig:efficiency_bubble} and~\ref{fig:tps_vs_power} suggest that variations in throughput lead to distinct power-performance operating regimes, which in turn directly influence the energy consumed per decision.

\begin{figure}[t]
  \centering
  \includegraphics[width=\columnwidth]{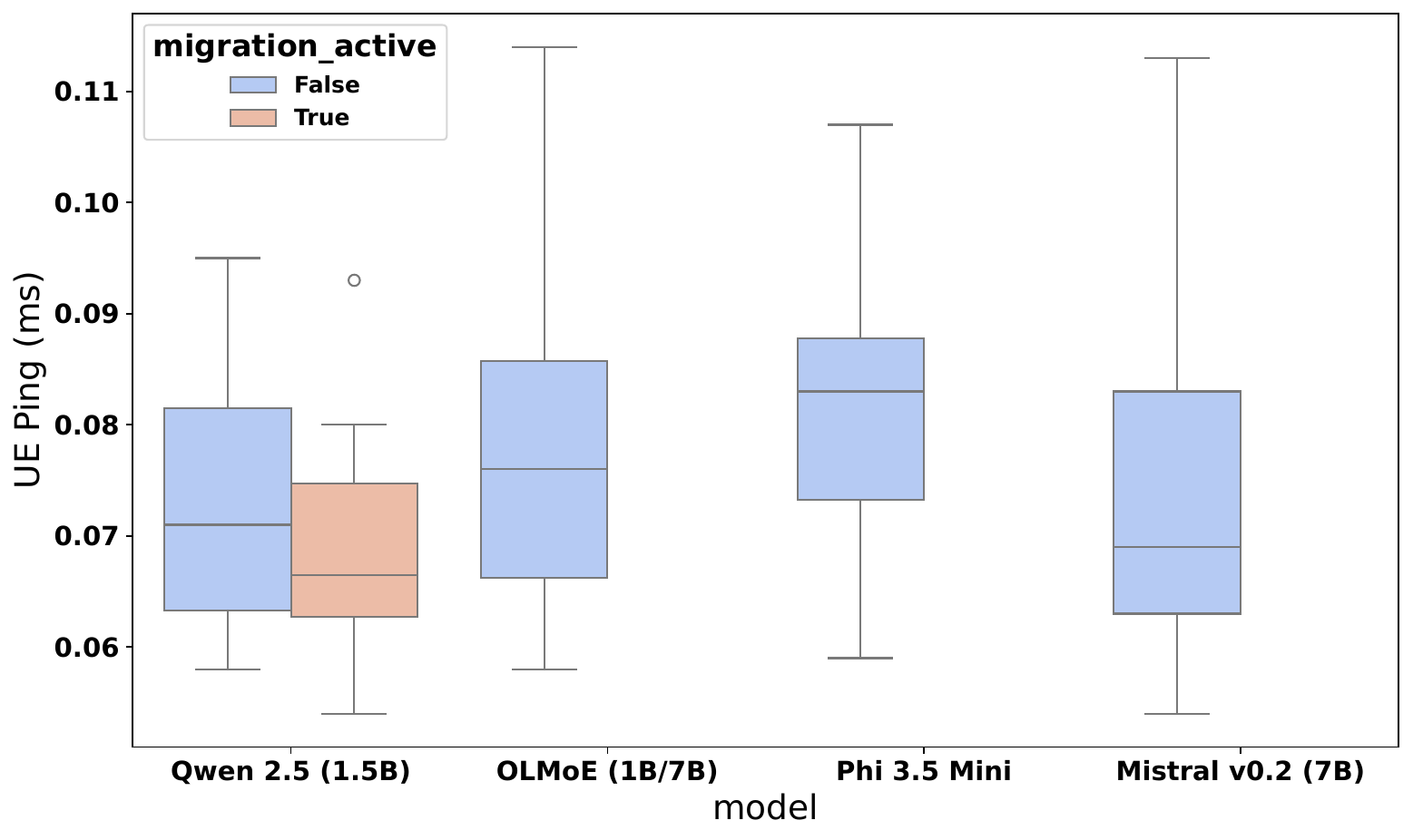}
  \caption{\ac{UE} latency grouped by model and migration state.}
  \label{fig:ue_ping_boxplot}
\end{figure}

\textbf{QoS impact at the \ac{UE}: latency and stability}. Figure~\ref{fig:ue_ping_boxplot} shows that \ac{UE} round-trip latency of the \ac{UE} remains within a narrow band (tens of milliseconds) across the models, indicating that the radio and data plane baseline dominates the mean \ac{UE} latency. However, the variability and tail behavior still revealed meaningful differences under stressed conditions. Qwen is the only model that yields both migration states in the dataset, enabling a direct within-model comparison. When migration occurs, the distribution shifts slightly downward. It tightens, suggesting that routing away from the stressed \ac{MEC} can reduce tail latency and improve stability, even if the average differences are modest.

\begin{figure}[t]
  \centering
  \includegraphics[width=\columnwidth]{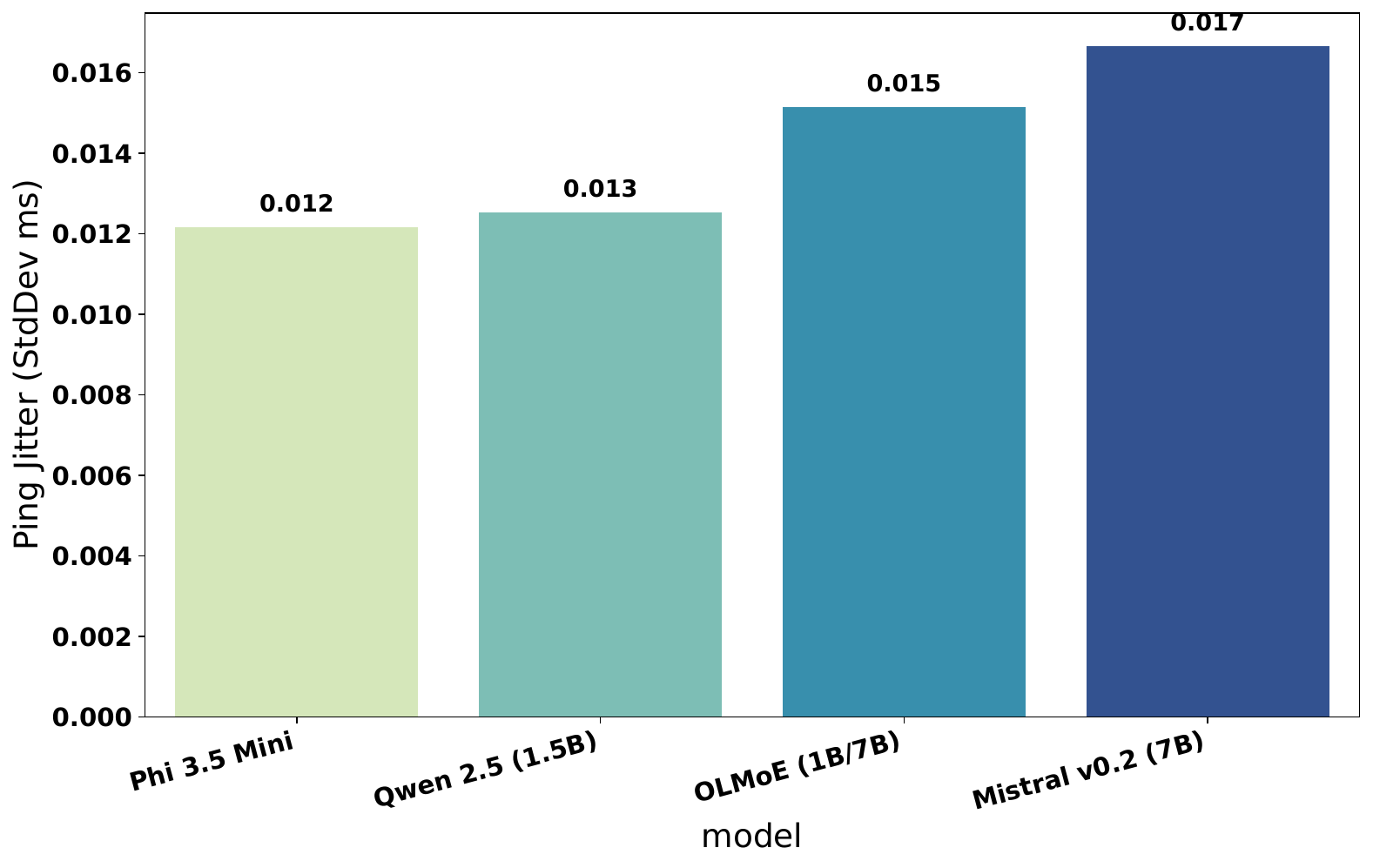}
  \caption{\ac{UE} latency standard deviation per model.}
  \label{fig:ue_jitter}
\end{figure}

The jitter proxy shown in Figure~\ref{fig:ue_jitter} reinforces this stability. Phi and Qwen showed the lowest dispersion (approximately 0.012--0.013), OLMoE was slightly higher (0.015), and Mistral exhibited the highest jitter (0.017). Notably, this ordering does not strictly follow the inference speed, indicating that \ac{UE} stability is influenced more by edge stress conditions and routing outcomes than by the agent runtime alone.

\begin{figure}[t]
  \centering
  \includegraphics[width=\columnwidth]{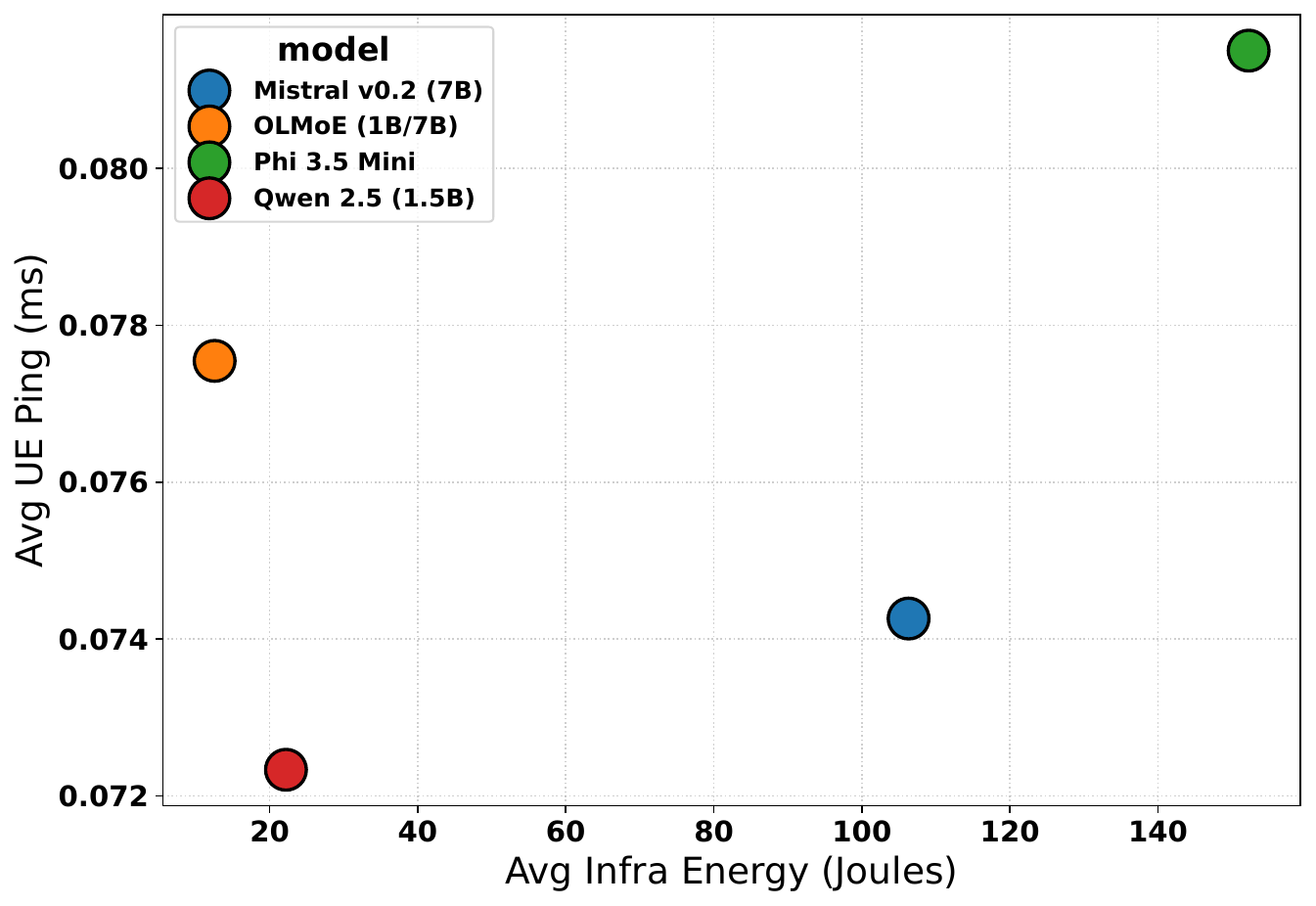}
  \caption{Average infrastructure energy versus average \ac{UE} latency per model.}
  \label{fig:pareto}
\end{figure}

\textbf{Joint energy-QoS trade-off}. Figure~\ref{fig:pareto} illustrates the operating points of each model concerning the average infrastructure energy and average \ac{UE} latency. Qwen is positioned in the advantageous lower-left region, offering the lowest average infrastructure energy and the lowest average \ac{UE} latency, thereby achieving the best joint operating point in this experiment. Although OLMoE further reduces energy consumption, it does so at a slightly higher average \ac{UE} latency than Qwen. Phi is located in the upper-right region, indicating a less favorable operating point, whereas Mistral occupies an intermediate position but is not Pareto-optimal, given Qwen's superior standing.

\begin{figure}[t]
  \centering
  \includegraphics[width=\columnwidth]{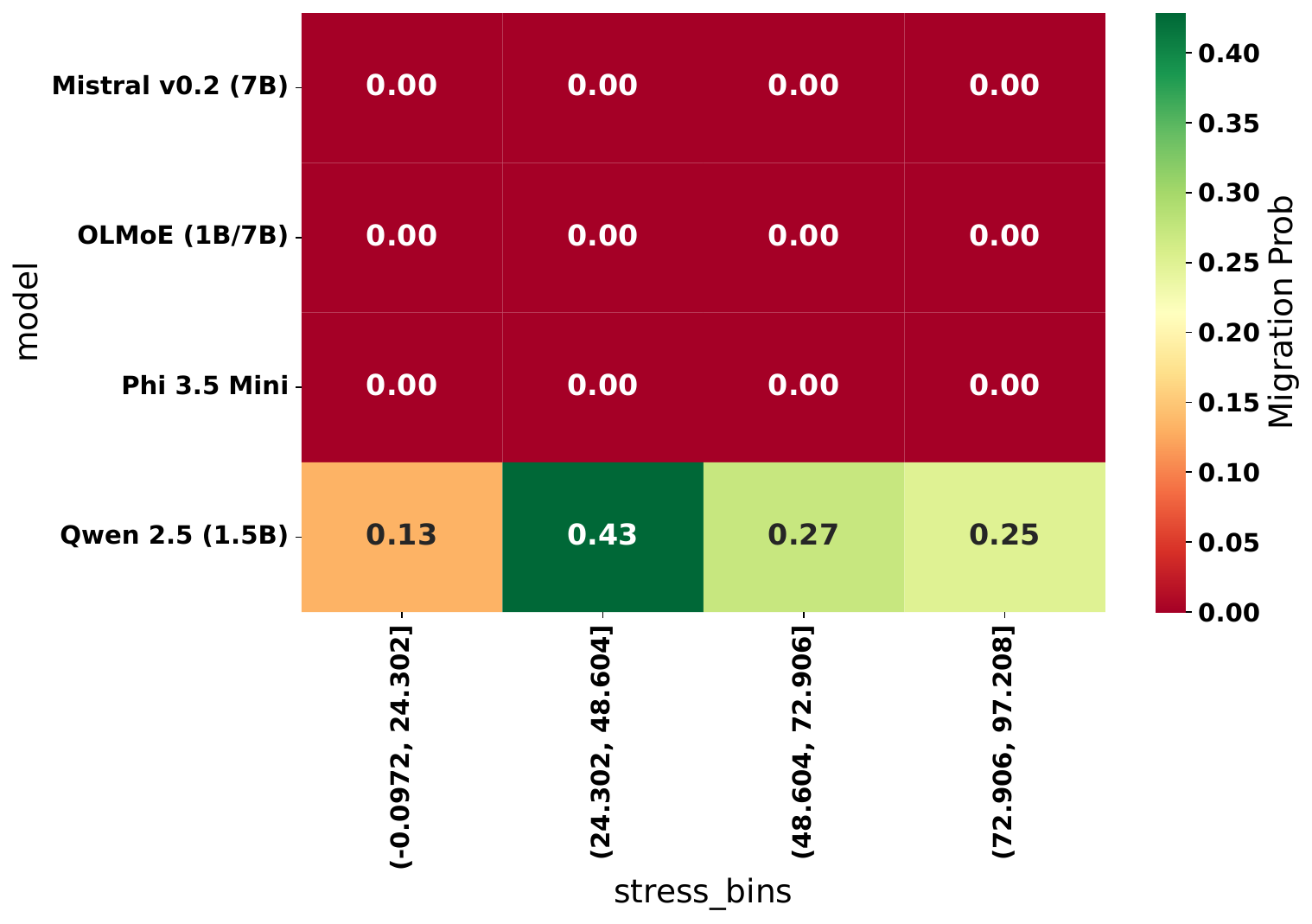}
  \caption{Migration probability by stressed \ac{MEC}~2 power bin.}
  \label{fig:policy_heatmap}
\end{figure}

\textbf{Policy execution under stress: migration behavior}. Performance metrics alone do not reveal whether an agent follows the intended policy. Figure~\ref{fig:policy_heatmap} shows the probability of triggering migration as a function of stressed \ac{MEC}~2 power, discretized into bins. Only Qwen triggered migrations with non-zero probability across bins (peaking at approximately 0.43), whereas all other models remained at zero. This outcome indicates that, under the same tool-calling interface and prompts, Qwen is the only model that reliably translates the energy threshold policy into an actionable \ac{UPF} update. Consequently, part of the energy and \ac{QoS} differences observed in earlier figures can be attributed not only to runtime but also to decision compliance.

\begin{figure}[t]
    \centering
    \includegraphics[width=\columnwidth]{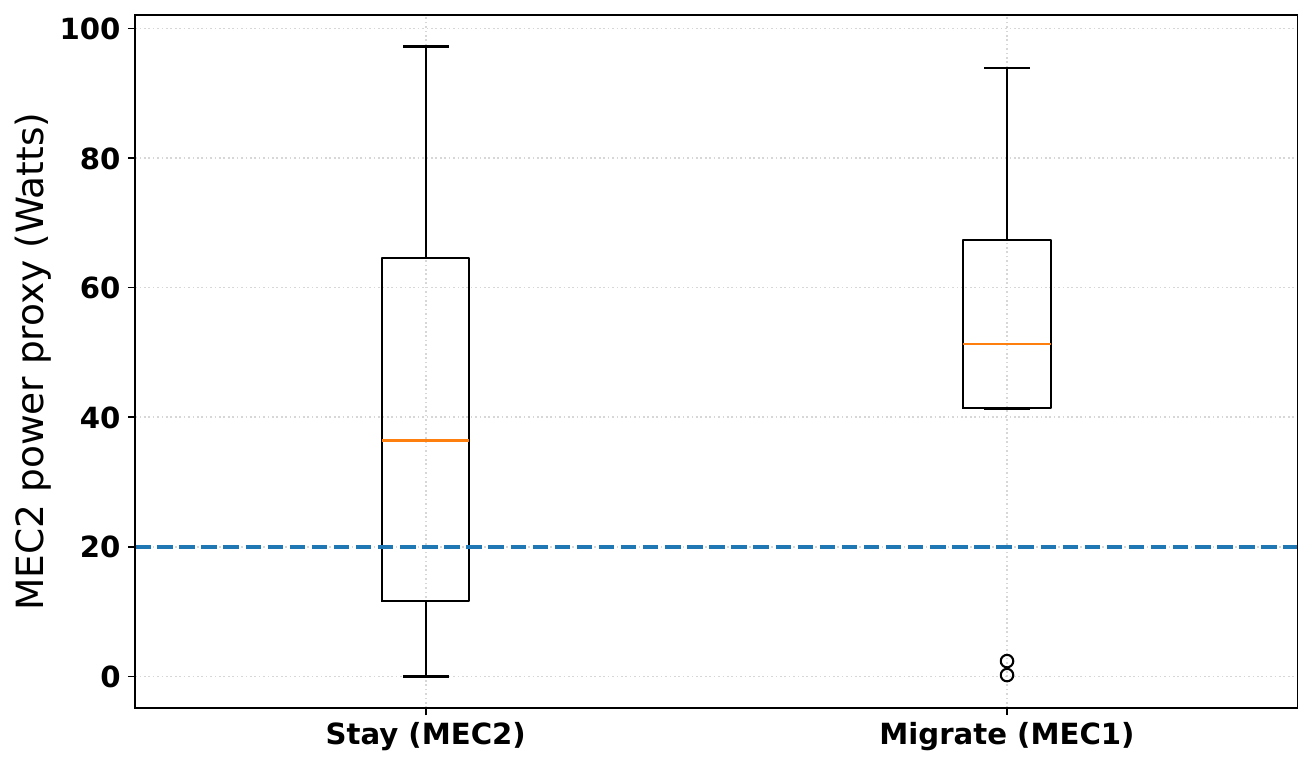}
    \caption{Qwen migration selectivity.}
    \label{fig:qwen_p2_by_action_precision}
\end{figure}

Together, these results show that telemetry-grounded tool use can be translated into verifiable \ac{UPF} actions, but decision compliance is model-dependent, making lightweight models with reliable tool/action alignment critical for sustainability-first closed-loop operation.

To complement the Figure.~\ref{fig:policy_heatmap}, Figure ~\ref{fig:qwen_p2_by_action_precision} provides a policy-grounded view of when AGORA uses Qwen to evaluate migration. The boxplots compare the \ac{MEC}2 power proxy ($P_2$) observed during decisions that \emph{kept} traffic on \ac{MEC}2 (Stay) versus decisions that \emph{migrated} the \ac{UPF} target to \ac{MEC}1 (Migrate), with the dashed line indicating the threshold $\theta = 20$\,W in Eq.~(\ref{eq:3}). The dashed line shows a power limit of 20\,W. Two main points are clear from this study. First, Qwen tends to divert traffic when \ac{MEC}2 uses more power. The power used during \emph{Migrate} was higher than that during \emph{Stay}. Second, the system is selective in nature. Although it correctly moved traffic only 28.57\% of the time when needed (true positive rate), it was accurate 80.00\% of the time when it did move traffic (positive predictive value). It also keeps the number of incorrect moves low at 15.38\% (false positive rate). This shows that the system avoids unnecessary changes but still acts when required.

\subsection{Workload context and GPU operational behavior}
\label{subsec:results_workload_gpu}

\begin{figure}[t]
  \centering
  \includegraphics[width=\columnwidth]{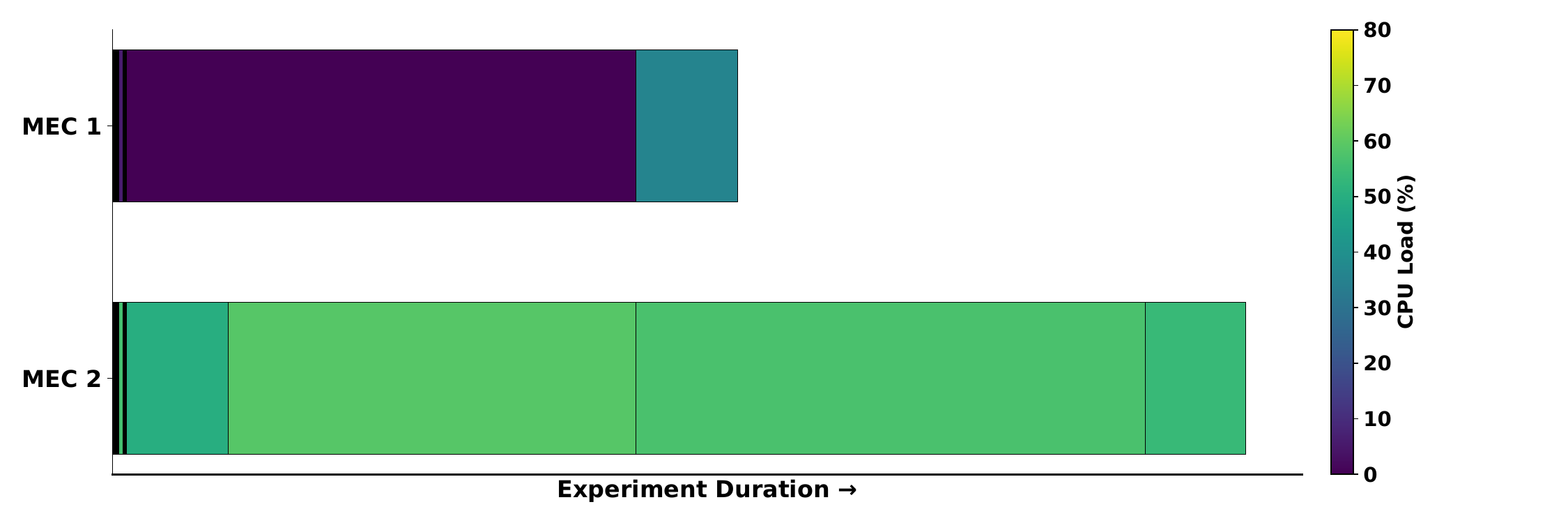}
  \caption{CPU load timeline applied to \ac{MEC}~1 and \ac{MEC}~2.}
  \label{fig:stress_timeline}
\end{figure}

\textbf{Stressor timeline and load asymmetry across MECs}. Figure~\ref{fig:stress_timeline} summarizes the injected CPU workload over time. The experiment enforces an asymmetric regime in which \ac{MEC}~2 sustains long intervals of elevated load, whereas \ac{MEC}~1 remains predominantly near idle with only brief activity. This asymmetry is essential for interpreting policy behavior because it ensures that failure to migrate keeps the \ac{UPF} target on a persistently stressed edge node, increasing the likelihood of revealing energy penalties and \ac{QoS} degradation. Conversely, any observed stability improvements when migration occurs can be attributed to routing decisions made under measurable stress rather than uniform background conditions.

\begin{figure}[t]
  \centering
  \includegraphics[width=\columnwidth]{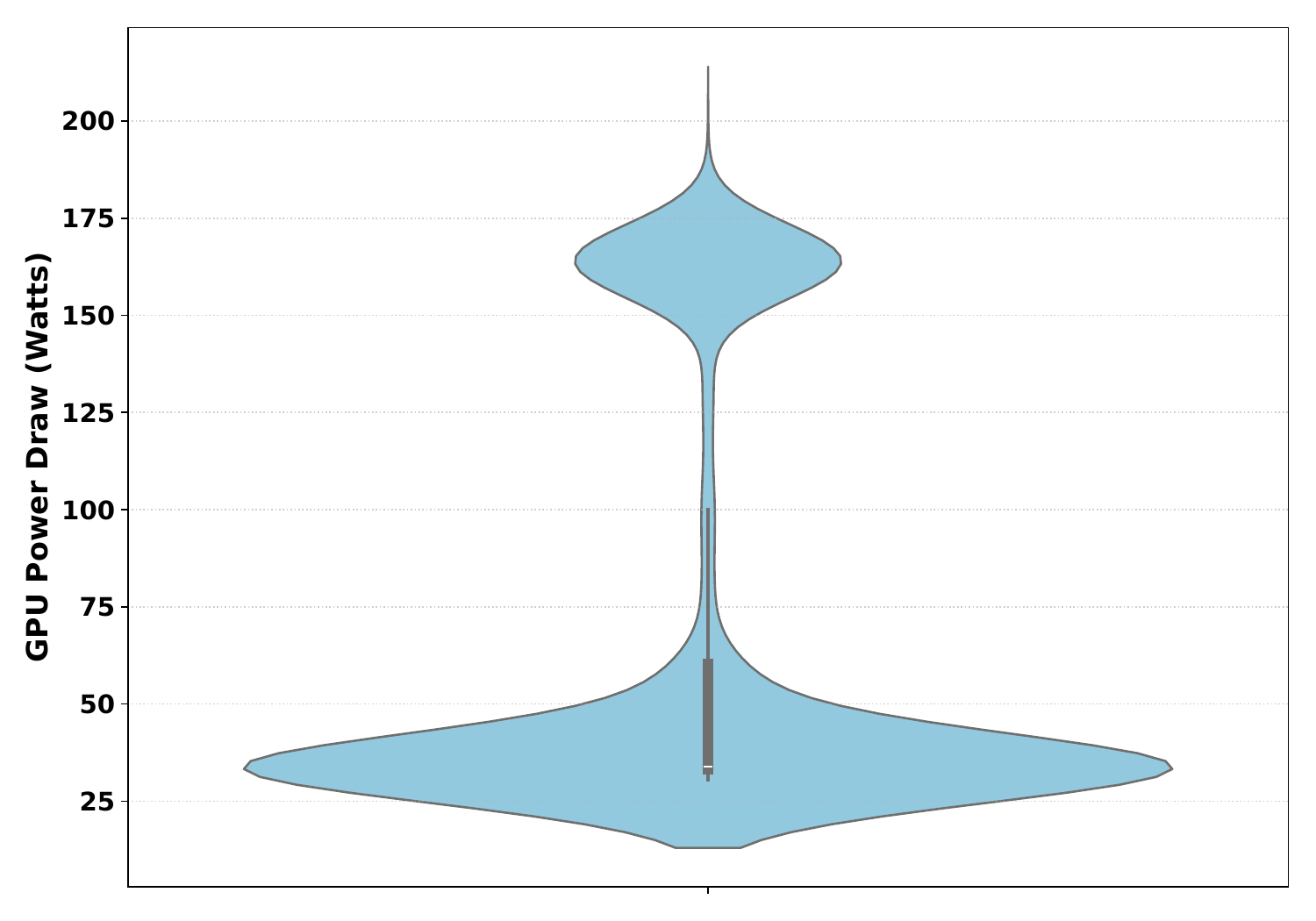}
  \caption{GPU power draw distribution over the experiment.}
  \label{fig:gpu_power_violin}
\end{figure}

\textbf{GPU power dynamics: bimodality and bursty execution}. Figure~\ref{fig:gpu_power_violin} shows the distribution of GPU power draw during the experiment. The shape is strongly bimodal, with one dominant low-power band centered at 30-40~W and a second high-power band centered at 160-175~W. This pattern is consistent with bursty inference: the system alternates between an idle or lightly utilized state and a saturated state. An important implication is that the average power can obscure operational reality: substantial time may be spent in low-power mode while still reaching high-power peaks during inference bursts, which is vital for thermal constraints and energy-aware orchestration policies.

\begin{figure}[t]
  \centering
  \includegraphics[width=\columnwidth]{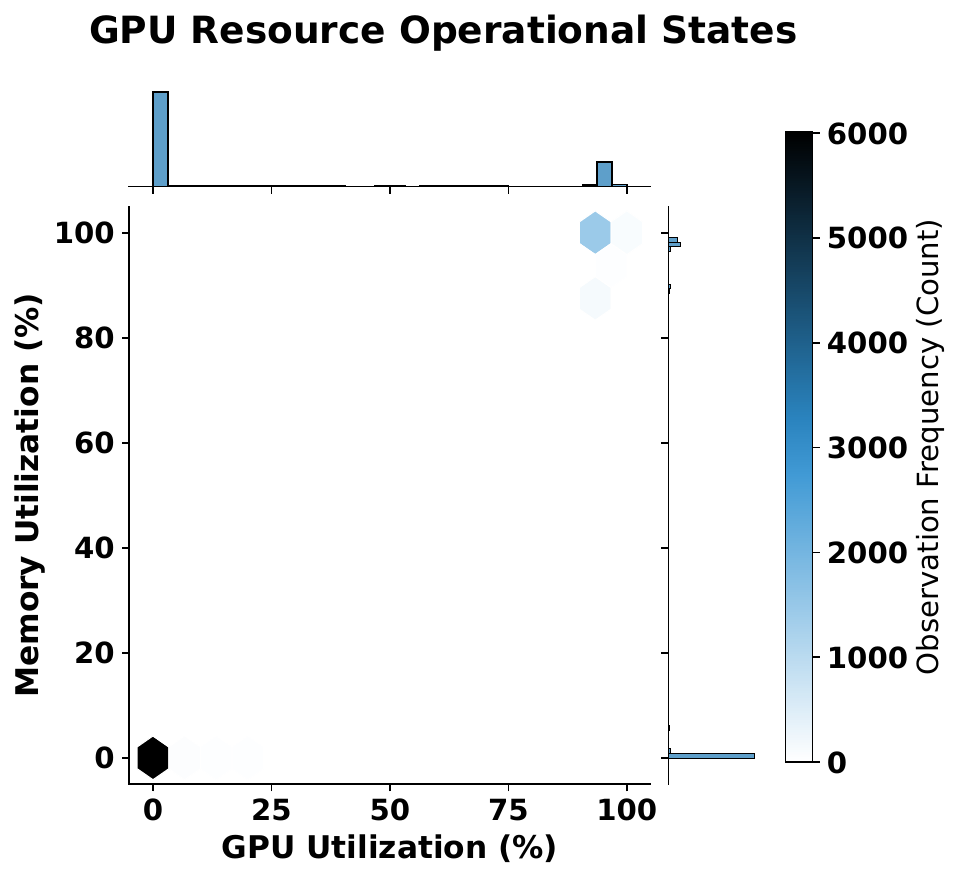}
  \caption{Density of GPU utilization versus GPU memory utilization.}
  \label{fig:gpu_hexbin}
\end{figure}

\textbf{GPU operational states: utilization versus memory utilization}. Figure~\ref{fig:gpu_hexbin} complements the power distribution by mapping the GPU utilization against the memory utilization with a density representation. The observations were concentrated near the bottom-left corner (approximately 0\% utilization and 0\% memory utilization) and formed a smaller cluster near the high-utilization region. The two clusters indicate that the accelerator frequently remained quiescent and transitioned to a high-activity state during inference. This separation supports scheduling strategies that reduce unnecessary high-power residency, for example, by consolidating requests and minimizing warm, underutilized periods.

\begin{figure}[t]
  \centering
  \includegraphics[width=\columnwidth]{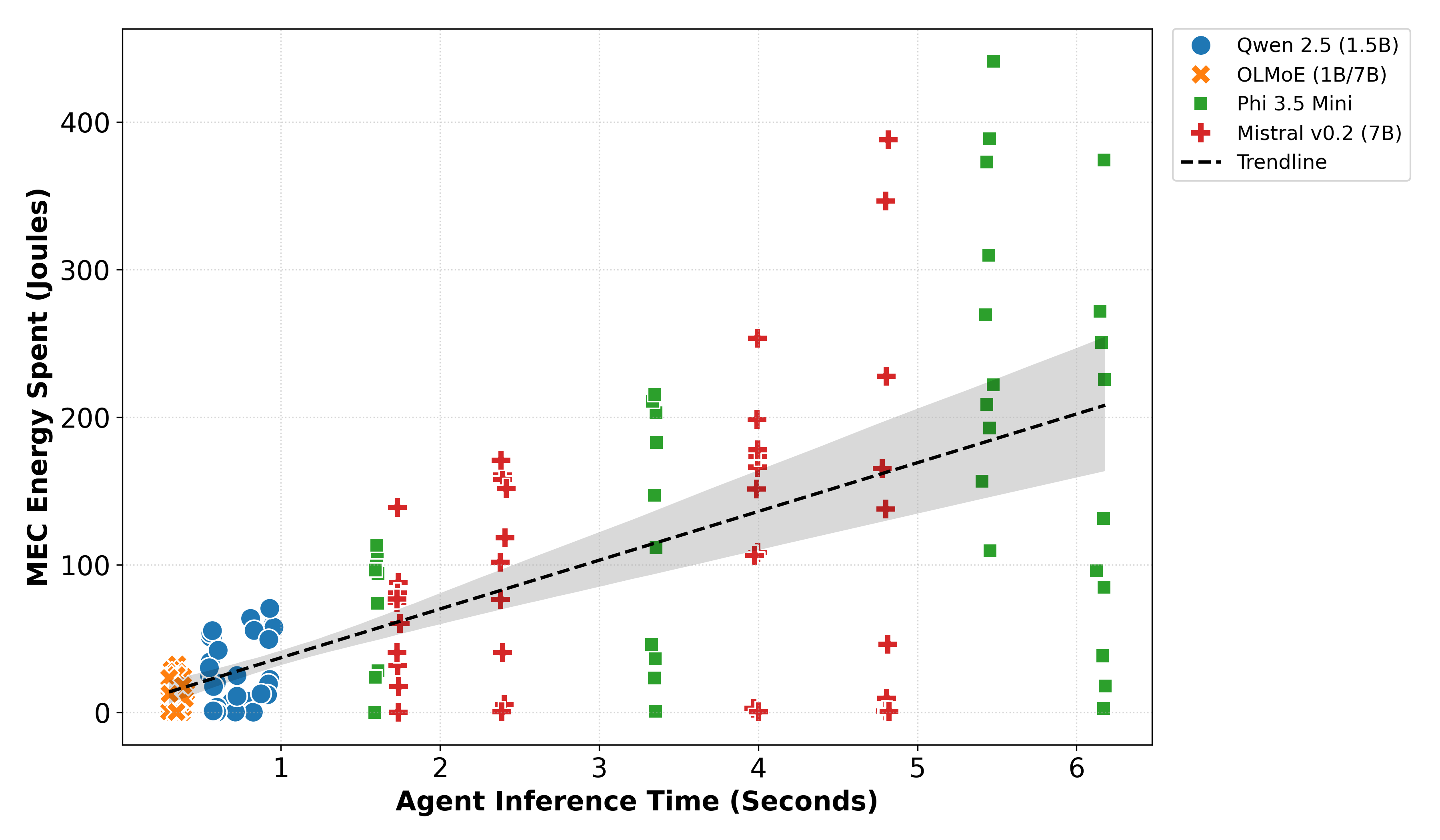}
  \caption{Inference time versus active \ac{MEC} energy.}
  \label{fig:latency_energy_corr}
\end{figure}

\textbf{Latency-energy coupling during decision loops}. Finally, Figure~\ref{fig:latency_energy_corr} directly relates the AGORA inference time to the corresponding active \ac{MEC} energy. The positive trend indicates that longer decision loops translate into higher infrastructure-energy expenditures. Model clusters further clarify this coupling: lightweight models occupy the low-latency/low-energy region, whereas larger models span a wider range of latencies and consume higher energy. The dispersion among larger models suggests sensitivity to run-to-run variability, such as caching behavior, tool invocation timing, and transient system conditions. Overall, this relationship supports the earlier conclusion that model choice impacts both responsiveness and energy footprint in closed-loop tool-augmented orchestration.

\section{Concluding Remarks}\label{sec:concluding_remarks}

In this paper, we propose an agentic orchestration architecture for energy-efficient Beyond-5G operation that enables sustainability-driven control of the mobile network data plane. Most importantly, AGORA operationalizes sustainability intents through telemetry-grounded, verifiable \ac{UPF} actuation, partially closing the gap between human sustainability goals and executable \ac{B5G} data-plane control. While prior work advances intent-based and zero-touch management, it rarely operationalizes sustainability objectives as first-class constraints at the \ac{5G} core, particularly at the \ac{UPF}.

Our approach integrates local \acp{LLM} into a telemetry-grounded, tool-augmented control loop that translates natural-language intents into executable function calls and \ac{UPF} routing actions. Experimental results show a strong latency-energy coupling in tool-driven decision loops and indicate that compact models can achieve low energy footprints while still enabling correct policy execution, including non-zero migration behavior under stressed \ac{MEC} conditions.

In future work, we will extend the architecture along three directions. First, we will broaden the model space by evaluating additional open \acp{LLM} and multi-agent configurations under larger intent suites and more diverse traffic patterns, including multi-slice and multi-tenant scenarios. Second, we will generalize tool integration beyond ad hoc wrappers by adopting standardized interfaces such as \ac{MCP} and Coral Protocol, enabling safer tool invocation, richer observability connectors, and portable agent deployments across platforms. Third, we will strengthen sustainability awareness by incorporating carbon intensity signals, risk-aware policies, and longer-horizon objectives, and by validating the approach on larger-scale testbeds with heterogeneous \acp{MEC} and real-world workload traces.

\bibliographystyle{IEEEtran}
\bibliography{references}

\begin{IEEEbiography}[{\includegraphics[width=1in,height=1.25in,clip,keepaspectratio]{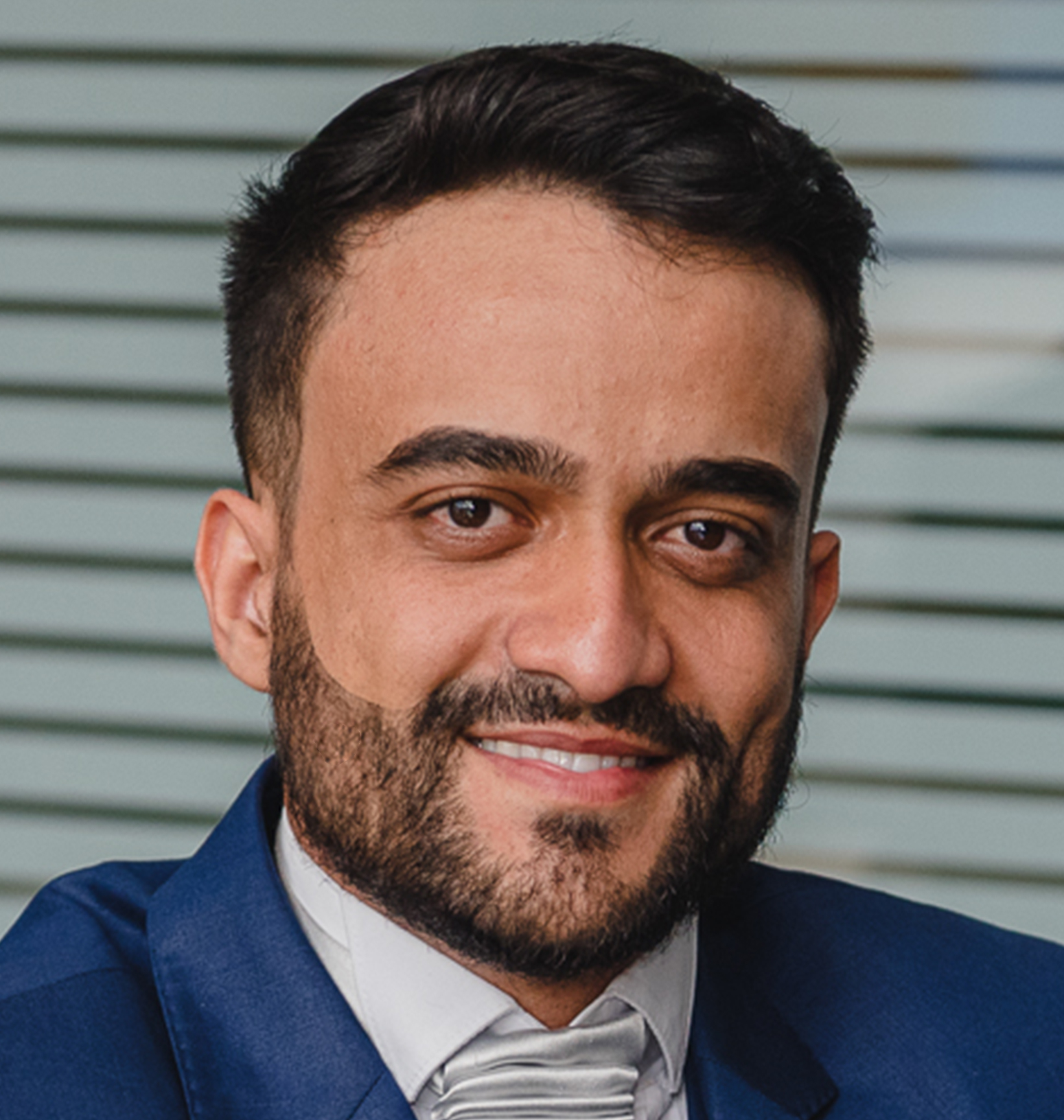}}]{Rodrigo Moreira} received his Ph.D. from the University of Uberlândia (UFU) in 2021. He is currently a Professor at the Federal University of Viçosa, where he also obtained his B.S. degree in 2014. He earned his M.Sc. degree from the Federal University of Uberlândia, Brazil, in 2017. He has several papers published and has presented at conferences. His research interests include the future of the internet, quality of service, cloud computing, network function virtualization, software-defined networking, computational intelligence, and edge computing.
\end{IEEEbiography}

\begin{IEEEbiography}[{\includegraphics[width=1in,height=1.25in,clip,keepaspectratio]{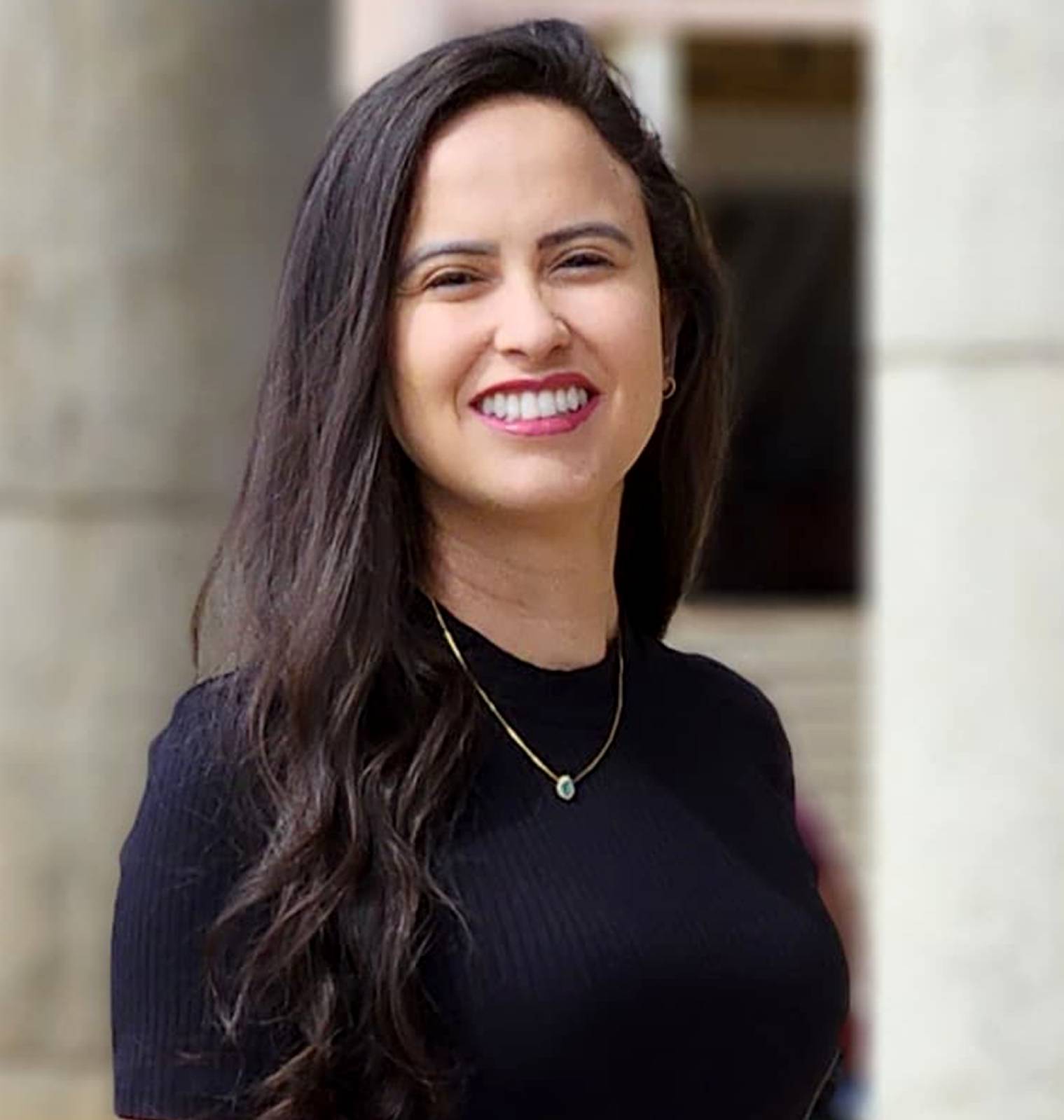}}]{Larissa Ferreira Rodrigues Moreira} received a Ph.D. in Computer Science (2024) from the Federal University of Uberlândia, Brazil. She received her B.Sc. degree in Computer Information Systems (2016) and M.Sc. degree in Computer Science (2018), both from Federal University of Viçosa, Brazil. Larissa is currently a Professor at Federal University of Viçosa. Her research interests include Artificial Intelligence, Computer Vision, and Image Processing.
\end{IEEEbiography}

\begin{IEEEbiography}[{\includegraphics[width=1in,height=1.25in,clip,keepaspectratio]{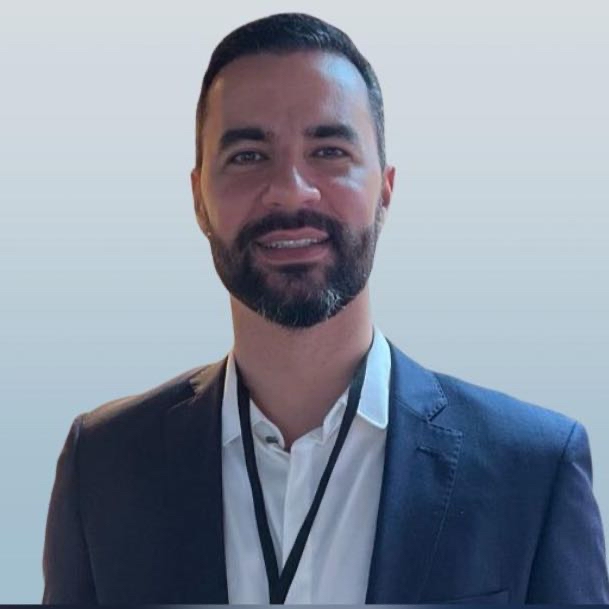}}]{Maycon Leone Maciel Peixoto} is an Associate Professor at the Institute of Computing of the Federal University of Bahia (UFBA). He holds a Ph.D. in Computer Science from the University of São Paulo (USP), Brazil, in 2012, and a Master's Degree in Computer Science from the same university, in 2008. He conducted a Postdoctoral (2020) at the University of Campinas (UNICAMP) and was a visiting scholar at the University of Toronto, Canada (2023/2024). He serves on the leading network and distributed system committees. Over the years, he has participated and coordinated several projects funded by public agencies and the private sector, with research interests spanning edge–cloud continuum, artificial intelligence and federated learning, vehicular and IoT networks, quantum computing, and data-driven digital transformation.
\end{IEEEbiography}

\begin{IEEEbiography}[{\includegraphics[width=1in,height=1.25in,clip,keepaspectratio]{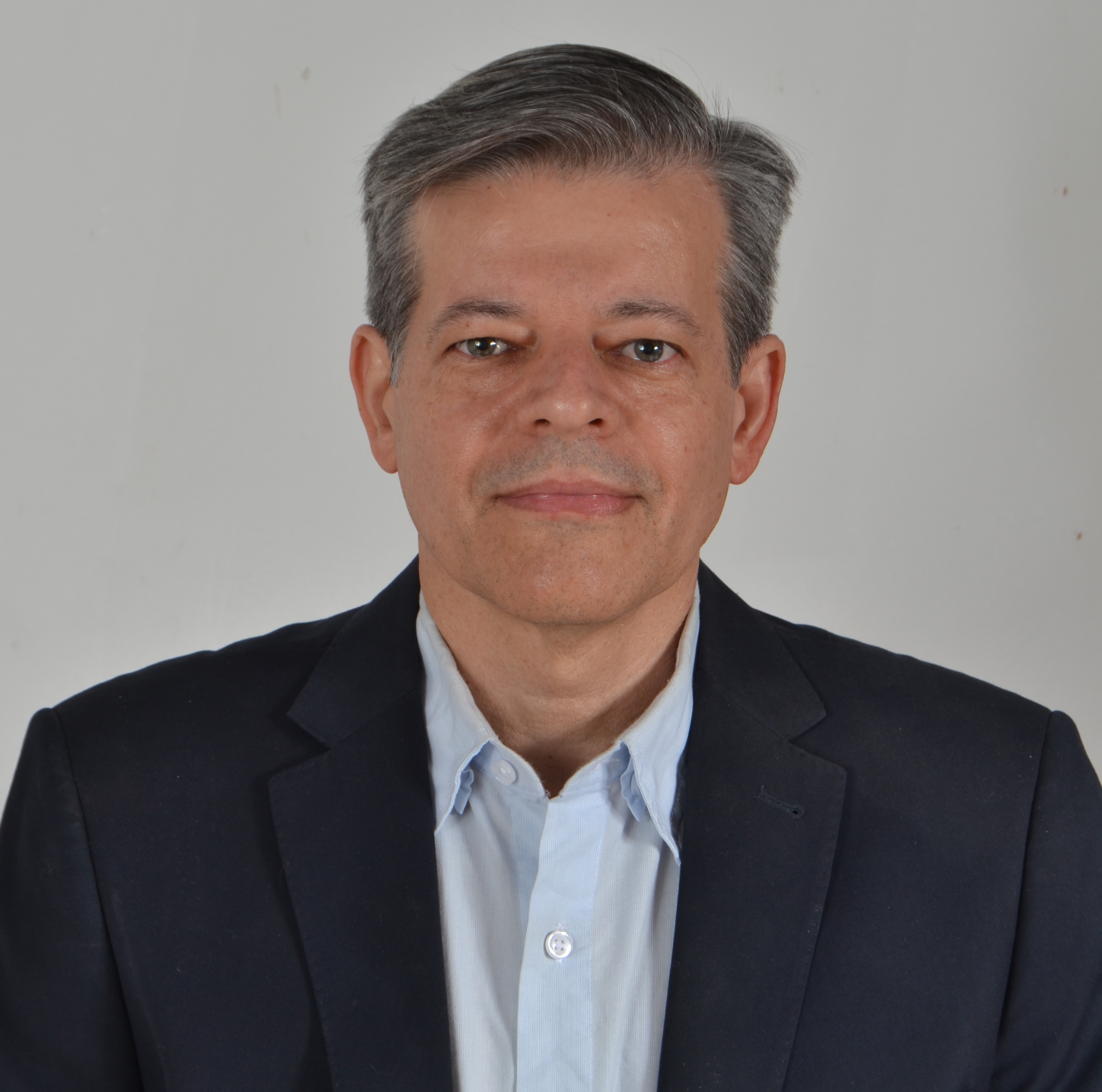}}]{Flávio de Oliveira Silva (Senior Member, IEEE)} received the Ph.D. from the University of São Paulo (USP) in 2013. He is a Professor in the Department of Informatics (DI) at the School of Engineering of the University of Minho in Braga, Portugal, and a researcher with the ALGORITMI Centre. He has published and presented several papers at conferences worldwide. His research interests include future networks, the IoT, network softwarization (SDN and NFV), future intelligent applications and systems, cloud computing, and software-based innovation. He is a member of IEEE, ACM, and SBC. He is a reviewer of several journals and a member of the TPC at several IEEE conferences.
\end{IEEEbiography}

\EOD

\end{document}